\documentclass[12pt]{iopart}

\usepackage{epsfig}
\usepackage{graphicx}
\usepackage{float}
\usepackage{xcolor}
\usepackage{mwe}

\begin{document}

\title[Anderson localization of memory random walks]{Anderson-like localization transition of random walks with resetting}

\author{Denis Boyer$^1$, Andrea Falc\'on-Cort\'es$^1$, Luca Giuggioli$^2$ and Satya N. Majumdar$^3$ }
\address{$^1$ Instituto de F\'\i sica, Universidad Nacional Aut\'onoma de M\'exico, Mexico City 04510, Mexico 

$^2$ Bristol Centre for Complexity Sciences, Department of Engineering Mathematics and School of Biological Sciences, University of Bristol, Bristol, BS8 1UB, UK 

$^3$ LPTMS, CNRS, Univ. Paris-Sud, Universit\'e Paris-Saclay, 91405 Orsay, France}
\ead{boyer@fisica.unam.mx}

\begin{abstract}
We study several lattice random walk models with stochastic resetting to previously visited sites which exhibit a phase transition between an anomalous diffusive regime and a localization regime where diffusion is suppressed. The localized phase settles above a critical resetting rate, or rate of memory use, and the probability density asymptotically adopts in this regime a non-equilibrium steady state similar to that of the well known problem of diffusion with resetting to the origin. The transition occurs because of the presence of a single impurity site where the resetting rate is lower than on other sites, and around which the walker spontaneously localizes. Near criticality, the localization length diverges with a critical exponent that falls in the same class as the self-consistent theory of Anderson localization of waves in random media. The critical dimensions are also the same in both problems. Our study provides analytically tractable examples of localization transitions in path-dependent, reinforced stochastic processes, which can be also useful for understanding spatial learning by living organisms.
\end{abstract}

\vspace{2pc}
\maketitle

\section{Introduction}

The study of stochastic processes with resetting has received an increasing attention in recent years. A paradigmatic example is given by a Brownian particle whose position is reset with a constant rate back to the origin, from which motion starts anew \cite{evans2011diffusion,evans2011diffusionb}. Owing to the fact that resetting events break detailed balance, the probability density of the particle position develops a non-equilibrium steady state (NESS), instead of following the standard Gaussian distribution. The NESSs of diffusive systems with resetting have been characterized in arbitrary spatial dimension \cite{evans2014diffusion}, for multiplicative processes \cite{manrubia1999stochastic}, continuous time random walks \cite{mendez2016characterization}, or in presence of absorbing boundaries and partially absorbing traps \cite{giuggioli2018comparison}. Furthermore, resetting is interesting for applications to random search problems \cite{evans2011diffusion,evans2011diffusionb}, or problems that require the completion of a random computing task \cite{reuveni2016optimal}, since the mean time needed to reach a target state for the first time by a process with resetting is finite and may be minimized with respect to the resetting rate. In addition, the relative fluctuations of the first passage times are equal to unity at the optimal rate and this property is universal \cite{reuveni2016optimal,pal2017first,belan2018restart}. 

A number of systems subject to resetting have been further studied, including diffusing particles with drift \cite{montero2013monotonic} and in external potentials \cite{pal2015diffusion}, active particles with run-and-tumble motion \cite{evans2018run}, L\'evy flights \cite{kusmierz2014first,kusmierz2015optimal}, processes characterized by a time-dependent resetting rate \cite{pal2016diffusion,kusmierz2018robust} or a time-dependent diffusion coefficient \cite{bodrova2018non}. Extensions to cases with non-exponential distributions between resetting events have allowed the emergence of a general understanding of the steady states and first passage properties in these systems \cite{pal2016diffusion,eule2016non,nagar2016diffusion,chechkin2018random}. The existence of phase transitions specific to resetting processes have also been discovered, such as a discontinuity in the optimal search time of L\'evy flights with restart \cite{kusmierz2014first,campos2015phase} and a transition in the temporal relaxation toward the steady state \cite{majumdar2015dynamical}. The effects of resetting on interacting systems with many degrees of freedom, such as fluctuating interfaces \cite{gupta2014fluctuating} or predator-prey populations \cite{mercado2018lotka} have also been studied. 

A class of processes of particular interest are those that do not restart always from the same point, like the initial state, but from any previously \emph{visited} state. Consider for instance a lattice random walker which relocates (at a given rate) to any site visited in the past according to a linear preferential rule: a site is chosen for a relocation with a probability proportional to the accumulated amount of time spent there by the walker since $t=0$, see {\it e.g.} Figure \ref{dynamic_m-g} below. The sites that are often visited are thus more likely to be visited again during a resetting event, which causes a spatial reinforcement. Such path-dependent processes are relevant to describe the tendency observed in many animals in the wild \cite{gautestad2004intrinsic,gautestad2006complex, merkle2014memory,boyer2014random} and humans \cite{song2010modelling} to frequently return to familiar places in their environment. 

Models with preferential resetting are among the few non-Markov random walks with long range memory for which a body of exact results have been derived. These walks do not exhibit NESSs but rather an anomalous diffusion where the mean square displacement (MSD) grows ultra-slowly with time, typically as $\ln t$ \cite{boyer2014random,boyer2016slow,boyer2017long}. Further studies have considered different forms of memory kernels, leading to a variety of behaviours ranging from non-equilibrium steady states to normal diffusion depending on the type of memory decay  \cite{ boyer2017long,boyer2014solvable,mailler2018random}. Central limit theorems and local limit theorems have been recently proven for these processes and their generalizations \cite{mailler2018random}.  Another study on resetting processes with memory has considered a random walk on a one-dimensional lattice with resetting to the rightmost visited position, showing that motion becomes ballistic with a speed and dispersion that depend on the resetting rate \cite{majumdar2015random}.

In a recent study \cite{falcon2017localization}, we have unveiled a remarkable phase transition in random walks with preferential resetting. Let us consider an infinite lattice containing one impurity site located at the origin (playing the role of a resource site or food), where the walker stays longer on average than on the other sites at each visit. As an emerging phenomenon of the reinforced dynamics, above a critical resetting rate the walker becomes localized around the impurity site and develops a NESS, similarly to a diffusing memory-less particle with stochastic resetting to the origin. Below the critical rate, however, the NESS is suppressed and slow diffusion takes place, basically as if the impurity was absent. These results indicate that very simple movement rules can actually allow the walker to {\it learn} about salient spatial features of an environment above a critical rate of memory use, similarly to a foraging animal adapting to its habitat and exploiting resources there. The transition is accompanied by a diverging localization length, characterized by a critical exponent that can be calculated within a decoupling approximation. The critical properties bear close relationships with the Anderson localization transition in a very different context \cite{anderson1958absence}.

Here we explore this problem further and present a general class of memory random walk models with resetting on heterogeneous lattices of arbitrary dimension. We analyse within a self-consistent theory the generic localization properties of three single-impurity models, including the model presented originally in \cite{falcon2017localization}. We review the main results on the localization transition in this system and elucidate the importance of heterogeneity in the resetting rate for the existence of a localization transition. We analyse the properties of the critical point at the critical dimension $D_c$, {\it i.e.}, the dimension below which localized states always exist. We also present analytic solutions in $1D$, numerical solutions in $2D$ and $3D$, and validate the results with Monte Carlo simulations.

\section{Model definitions and relation with previous work}\label{secmodel}

We start by introducing a quite general class of models of random walks with memory in inhomogeneous environments, that we further specify through a few illustrative examples. Let us consider an infinite $D$-dimensional cubic lattice and a walker with initial position $X_0=n_0$ at $t=0$. Each lattice site $n$ is characterized by two quenched probabilities $s_n$ and $r_n$, which are set before hand and do not change over time. Time is discrete and we denote $X_t$ as the position of the walker at time $t$. Assuming $X_t=n$, during the time step $t\rightarrow t+1$, the walker performs one of the 3 following actions:
\begin{itemize}
\item {\bf} With probability $s_n$, it stays on site $n$, that is $X_{t+1}=X_t$.

\item {\bf} With probability $r_n$, it resets to a visited site, that is $X_{t+1}=i$ where $i$ is one of the previously occupied site. The probability to choose a particular site $i$ is proportional to the total amount of time spent by the walker on that site during $[0,t]$.

\item {\bf} With probability $1-s_n-r_n$, the walker performs a non-zero random walk step $\ell_{t+1}$, drawn from a given symmetric distribution $p(\ell)$, namely, $X_{t+1}=X_{t}+\ell_{t+1}$.
\end{itemize}

\begin{figure}[H]
  \centerline{\includegraphics*[width=0.7\textwidth]{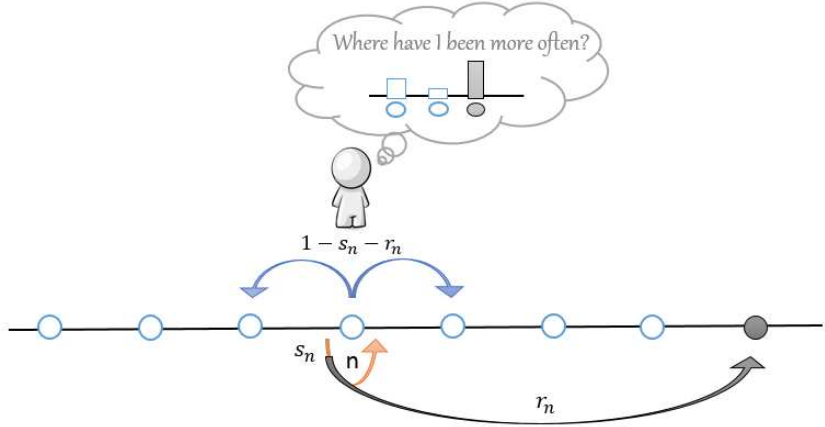}}
  \caption{Dynamics of the general model. With probability $s_n$ the walker stays at its current position $n$ one more time unit; with probability $r_n$ it resets to one of its previously visited sites, such site being chosen with a probability proportional to the total amount of time spent there so far. With the complementary probability $1-s_n-r_n$, the walker performs a random walk step; in this case the upper arrows indicate a nearest neighbour step.}
\label{dynamic_m-g}
\end{figure}

The diagram of Figure \ref{dynamic_m-g} displays these rules. Defining $P_n(t)$ as the probability that $X_t=n$, one can write the following master equation
\begin{eqnarray}\label{mastergen}
P_n(t+1)&=&\sum_{\ell}(1-s_{n-{\ell}}-r_{n-{\ell}})p({\ell})P_{n-{\ell}}(t)+s_nP_n(t)\nonumber\\
&&+\sum_m \frac{r_m}{t+1}\sum_{t'=0}^{t}{\rm Prob}[X_{t'}=n\ {\rm and}\ X_t=m],
\end{eqnarray}
in any dimension $D$.
The first two terms of the r.h.s. of Eq. (\ref{mastergen}) describe the diffusion and \lq\lq trapping" of the walker, whereas the last term asserts that the walker can return to site $n$ by a resetting event from any site $m$ occupied just before. In this case, one must count the number of time units that $n$ was occupied in the past and divide by the total elapsed time ($t+1$) to obtain the probability of resetting to $n$. An equivalent way of translating the preferential return rule (and which is used in the simulations below) is to choose a time in the past $t'$ uniformly in $[0,t]$, and to reset the walker to the position it occupied at that time \cite{boyer2014random,boyer2014solvable}. Later on, the $s_n$'s and $r_n$'s will be all equal, except potentially at the origin $n=0$.

The walker has memory of its whole history, since it remembers all its previous positions (or, equivalently, the time spent on each visited sites). In addition, its motion takes place in an inhomogeneous environment. We now discuss a few limiting cases of this model. 

{\it No resetting.} If $r_n=0$ $\forall n$, the problem reduces to a Markovian diffusion problem in a non homogeneous medium. If, in addition, $s_n=\gamma$ $\forall n$ with $\gamma$ a non zero constant, the so-called lazy random walk model is recovered \cite{shen2014lazy}. With $\gamma=0$, it reduces to the standard random walk with step distribution $p({\ell})$.

{\it Uniform resetting probability.} If $r_n=q>0$ $\forall n$, the memory term of Eq. (\ref{mastergen}) can be simplified thanks to the identity
\begin{equation}
  \sum_m {\rm Prob}[X_{t'}=n\ {\rm and}\ X_t=m]={\rm Prob}[X_{t'}=n]\equiv P_n(t').  
\end{equation}
If we also take $s_n=0$ everywhere, the master equation then reads
\begin{equation}\label{boyersolis}
P^{(0)}_n(t+1)=(1-q)\sum_{\ell}p({\ell})P^{(0)}_{n-{\ell}}(t)
+ \frac{q}{t+1}\sum_{t'=0}^{t}P^{(0)}_n(t').
\end{equation}
where the superscript $^{(0)}$ refers to the case $r_n=$ constant, $s_n=0$. Although non-local in time, Eq. (\ref{boyersolis}) solely involves the one-time distribution $P_n(t)$, instead of the one-time and two-time functions in Eq. (\ref{mastergen}). This property of (\ref{boyersolis}) is a direct consequence of the uniformity of $r_n$. Equation (\ref{boyersolis}) is completely homogeneous in space and exact asymptotic results were obtained in \cite{boyer2014random,boyer2016slow,mailler2018random}. Versions of the model in continuous space and continuous time were studied in \cite{boyer2017long,mailler2018random,campos2018recurrence}. The main results can be summarized as follows. The MSD grows as $\ln t$ asymptotically:
\begin{equation}\label{m2}
\langle (X_t-n_0)^2\rangle=\sum_{n}(n-n_0)^2P^{(0)}_n(t)\simeq \frac{1-q}{q}
\langle |{\ell}|^2\rangle\ln (qt),    
\end{equation}
a relation which holds true for any $0<q\le 1$. Hence, as soon as $q$ is non-zero, memory generates an ultra-slow growth of the MSD at large times. Despite such a strongly anomalous dynamics, the site occupation probability tends to a Gaussian asymptotically, similarly to a memory-less random walk,
\begin{equation}\label{slowgauss}
P^{(0)}_n(t)\rightarrow \frac{1}{\left(\sqrt{2\pi \langle (X_t-n_0)^2\rangle}\right)^{D}}
\exp\left[-\frac{(n-n_0)^2}{2\langle (X_t-n_0)^2\rangle}\right].
\end{equation}
In this expression, the variance does not grow as $t$ like in normal diffusion but follows the logarithmic law (\ref{m2}). The exact form of the distribution at all $t$ for nearest neighbour steps in $1D$ can be obtained in Fourier space, as exposed in the \ref{appa}. A rigorous proof of a central limit theorem for this problem and many generalizations was presented in \cite{mailler2018random}, with the help of a mapping to weighted random recursive trees.

In this work, we study the effects produced by adding a {\it single impurity} site in a system that otherwise would obey Eq. (\ref{boyersolis}). With just one impurity (that can affect either $s_n$, $r_n$ or both), the problem remains relatively simple and, still, exhibits behaviours markedly different from the homogeneous case and from normal diffusion. We consider three single-impurity models.

\subsection{Model I}



\begin{figure}[H]
  \begin{center}
     \includegraphics*[width=0.7\textwidth]{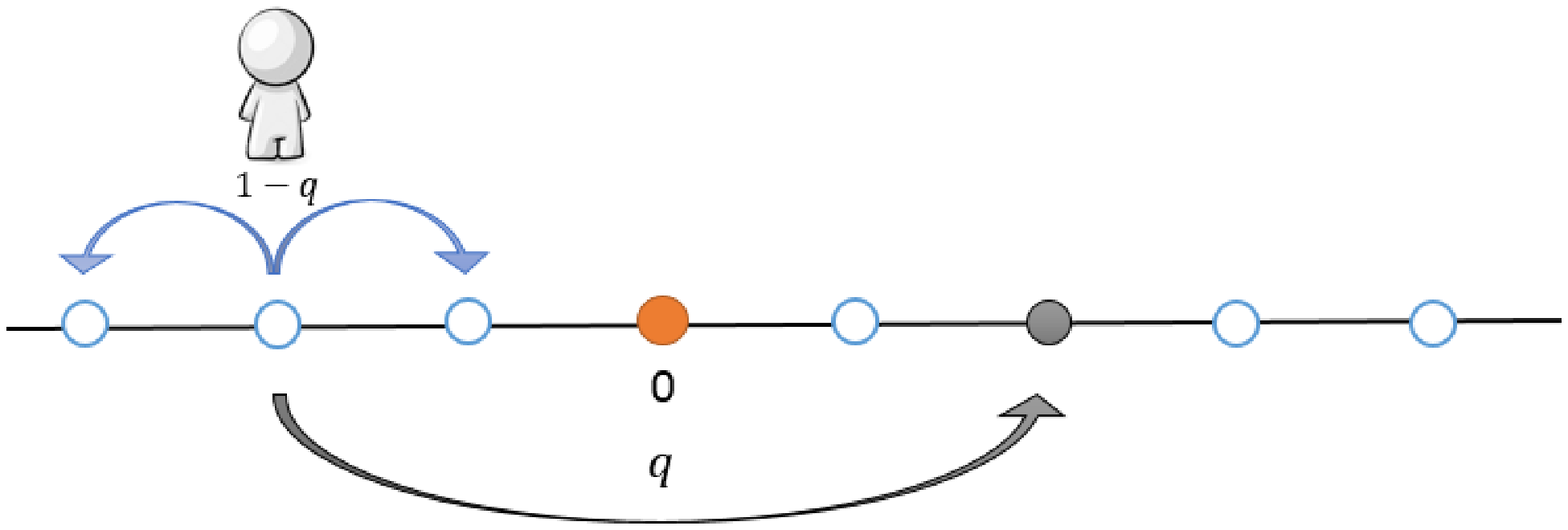}
     
     \vspace{0.5cm}
     
      \includegraphics*[width=0.7\textwidth]{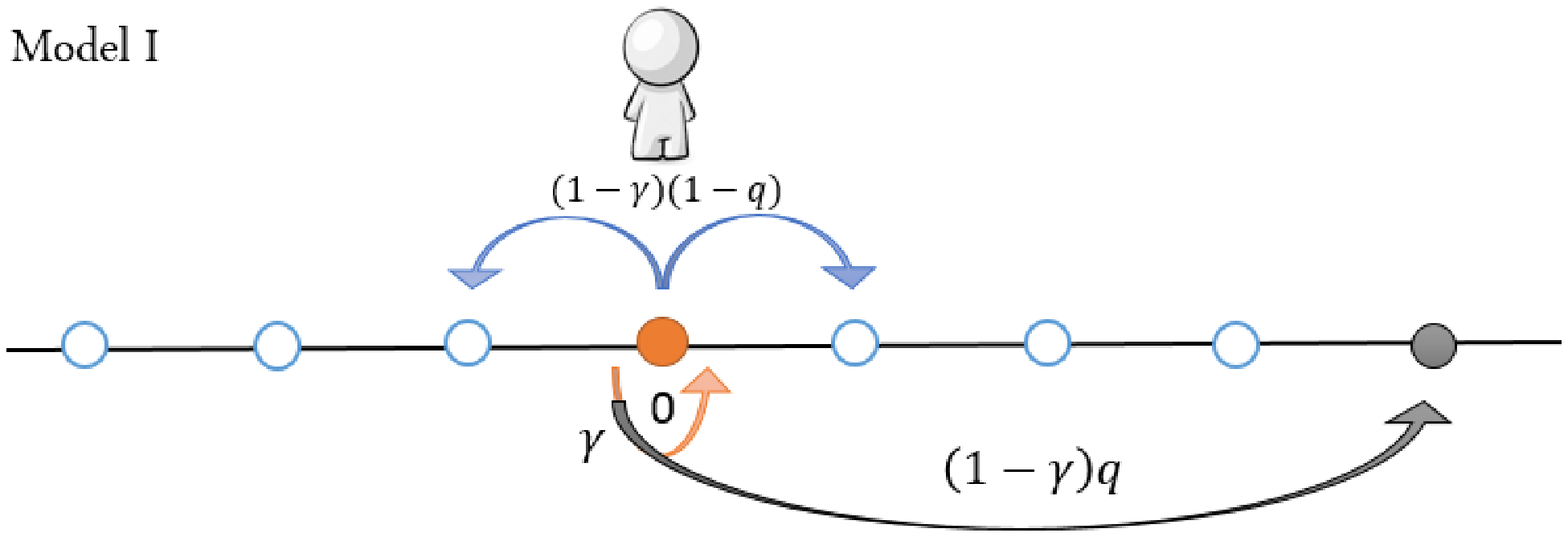}
  \caption{Model I. Top: When $X_t\ne0$, the walker takes a random step with probability $1-q$ and uses memory with preferential revisits with probability $q$.  Bottom: When $X_t=0$, the walker stays there one more time unit with probability $\gamma$ and moves according to the rules above with probability $1-\gamma$.}
\label{dynamic_m-im}
    \end{center}
\end{figure}

The first model mimics the behaviour of an animal moving in an environment containing one food site and was first presented in \cite{falcon2017localization}. In all the following, the impurity is located at the origin $n=0$. When the animal is not on the food site, as previously exposed, it either takes a random step, with probability $1-q$, or resets preferentially to a visited site, with probability $q$ (see Fig. \ref{dynamic_m-im}-Top). When the animal is located at the origin, it stays there with a probability $\gamma$ at the following time or moves with probability $1-\gamma$ according to the rules above (see Fig. \ref{dynamic_m-im}-Bottom). Therefore, when the walker is at the origin, it always considers the option of staying there one more time unit (with probability $\gamma$). The mean time spent at the origin during a visit can be thought of as the time for food consumption. This model is thus defined by the probabilities
\begin{eqnarray}
&& s_n=\gamma \delta_{n,0}\label{snmodelI} \\
&& r_n=q(1-\gamma \delta_{n,0}),\label{rnmodelI}
\end{eqnarray}
with $\delta_{n,0}$ the Kroneker symbol. Relation (\ref{rnmodelI}) asserts that the resetting rate is $q(1-\gamma)\le q$ on the impurity and $q$ elsewhere.

\subsection{Model II}
\begin{figure}[H]
  \centerline{\includegraphics*[width=0.7\textwidth]{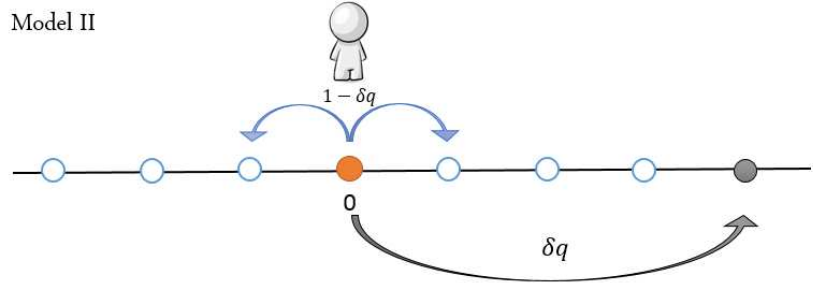}}
  \caption{Rules of Model II, when $X_t=0$. If $X_t\ne 0$, motion is like in Figure \ref{dynamic_m-im}-Top.}
\label{dynamic_m-ii}
\end{figure}

The second model is also inspired from foraging ecology, with a different response of the animal to the food site. 
On the food site, the animal has a vanishing staying probability (like on the other sites) but decides to favour random movement over resetting. In other words, realizing that there is food, the animal changes its behaviour to an increased local exploration, at the expense of using memory. This is modeled by decreasing the resetting rate at $n=0$ by a factor $\delta<1$, see Figure \ref{dynamic_m-ii}. Therefore,
\begin{eqnarray}
&& s_n=0\label{snmodelII}\\
&& r_n=q[1-(1-\delta) \delta_{n,0}],\label{rnmodelII}
\end{eqnarray}
(The Kroneker symbol $\delta_{n,0}$ must not be confused with the parameter $\delta$.) Relation (\ref{rnmodelII}) states that the resetting rate is $\delta q\le q$ on the impurity and $q$ elsewhere.

\subsection{Model III}

\begin{figure}[H]
  \centerline{\includegraphics*[width=0.7\textwidth]{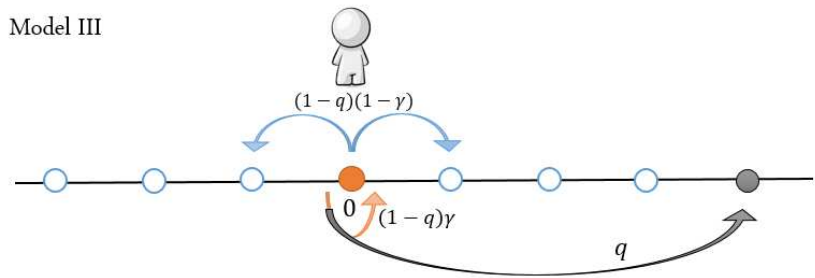}}
  \caption{Rules of Model III, when $X_t=0$. If $X_t\ne 0$, motion is like in Figure \ref{dynamic_m-im}-Top.}
\label{dynamic_m-iii}
\end{figure}

This model is similar to Model I, but with a uniform resetting rate. When the walker is not on the impurity, the transition probabilities are the same as in Models I and II. When the walker occupies the impurity, it keeps using the resetting mode with probability $q$. With the complementary probability $1-q$, it chooses among two possibilities: remaining on the impurity $[$with probability $\gamma]$ or moving at random to neighbouring site $[$with probability $1-\gamma]$,  as shown in Figure \ref{dynamic_m-iii}.

Like in Model I, setting $\gamma=0$ reduces to the homogeneous model, whereas with $\gamma>0$ the walker tends to spend some time at the special site at each visit. However, contrary to Model I, the walker remains on the impurity with probability $\gamma$ {\it only if it does not use its memory}. This subtle difference brings important consequences on the large time dynamics, as we will see. These rules can be recast as
\begin{eqnarray}
&& s_n=(1-q)\gamma \delta_{n,0}\label{snmodelIII}\\
&& r_n=q.\label{rnmodelIII}
\end{eqnarray}

\section{Results on Model I} 

\subsection{Summary of previous results}\label{sec:previous}

We recall the main results obtained on Model I in \cite{falcon2017localization}.
With the transition probabilities (\ref{snmodelI}) and (\ref{rnmodelI}), the master equation (\ref{mastergen}) takes the form:
    \begin{eqnarray}\label{mastermodelI}
     P_n(t+1) &=& (1-q)\sum\limits_{{\ell}} (1-\gamma \delta_{n-{\ell},0})p({\ell})P_{n-{\ell}}(t)+
     \gamma \delta_{n,0}P_n(t) \nonumber \\
              &+& \frac{q}{t+1}\sum\limits_{t'=0}^{t} \mbox{Prob}[X_{t'}=n\;\;\mbox{and}\;\;X_t\neq 0]\nonumber\\
              &+& \frac{q(1-\gamma)}{t+1}\sum\limits_{t'=0}^{t} \mbox{Prob}[X_{t'}=n\;\;\mbox{and}\;\;X_t= 0]
    \end{eqnarray}
Instead of trying to solve Eq. (\ref{mastermodelI}) at finite $t$, we take the limit $t\rightarrow\infty$ and seek non-equilibrium steady state (NESS) solutions, such that $\lim_{t\rightarrow\infty}P_n(t)=P_n\ne 0$.
We also use a de-correlation approximation, where $\mbox{Prob}[X_{t'}=n\;\;\mbox{and}\;\;X_t\neq 0]$ is replaced by $P_n(t')[1-P_0(t)]$, and $\mbox{Prob}[X_{t'}=n\;\;\mbox{and}\;\;X_t= 0]$ by $P_n(t')P_0(t)$, which is equivalent to assume that the positions $X_t$ and $X_{t'}$ occupied at different times by the walker become uncorrelated \cite{falcon2017localization}. Under this approximation the equation for the one-point function $P_n(t)$ closes and we can exactly solve it.
The solution obtained agrees very well at late time with numerical simulations, justifying a posteriori this approximation. Furthermore, the excellent agreement with simulations seem to indicate that the de-correlation approximation becomes asymptotically exact at late times, though we were not able to prove this rigorously. Hence, taking the limit $t\rightarrow\infty$ and $t'\rightarrow\infty$ in Eq. (\ref{mastermodelI}), the equation satisfied by the NESS, if it exists, becomes
\begin{equation}\label{nessmodelI}
 P_n = (1-q)\sum_{{\ell}} p({\ell})P_{n-{\ell}}+qP_n(1-\gamma P_0)
     + \gamma P_0[\delta_{n, 0}-(1-q)p(n)]. \nonumber
\end{equation}
We define the discrete Fourier transform of $P_n$ as
\begin{equation}
    \widehat{P}_k=\sum_{n}e^{-ikn}P_n,
\end{equation}
where the variable $k$ must be understood as the $D$ dimensional vector $\vec k=(k_1,...,k_D)$, and $kn$ as the dot product $\vec k\cdot \vec n$ with $\vec n=(n_1,...,n_D)$. Similarly, the variable ${\ell}$ for the displacements denotes the vector ${\vec\ell}=({\ell}_1,...,{\ell}_D)$. The transform of Eq. (\ref{nessmodelI}) yields
\begin{equation}\label{transPn}
 \widehat{P}_k=\frac{\gamma P_0[1-(1-q)\widehat{p}(\vec k)]}{(1-q)[1-\widehat{p}(\vec k)]+q\gamma P_0}.
\end{equation}
Notice that this solution is independent of the initial condition. In this expression, $\widehat{p}(\vec k)$ is the Fourier transform of the distribution of the random steps $p({{\vec{\ell}}})$, and $P_0$ is the asymptotic probability of occupying the impurity site. $P_0$ can be determined self-consistently by taking the inverse transform of $\widehat{P}_k$ evaluated at $n=0$:
\begin{equation}\label{SCgen}
    P_0=\frac{1}{(2\pi)^D}\int_{\cal B}d\vec k\ \widehat{P}_k.
\end{equation}
In $D$ dimension, ${\cal B}$ is the first Brillouin zone, defined by $-\pi\le k_i\le \pi$ for each component of $\vec k$. A self-consistent equation for $P_0$ is obtained by substituting Eq. (\ref{transPn}) into (\ref{SCgen}). Apart from the trivial solution $P_0=0$, which always exists, a non-trivial solution is predicted. It satisfies the transcendental equation
\begin{equation}\label{SCmodelI}
 \frac{1}{(2\pi)^D}\int_{\cal B} \frac{d\vec k}{(1-q)[1-\widehat{p}(\vec k)]+q\gamma P_0} = \frac{1-\gamma}{q\gamma(1-\gamma P_0)}.
\end{equation}
Fixing $\gamma$, Eq. (\ref{SCmodelI}) does not have a solution for all values of $q$ in general. To see this, let us set $P_0$ to its minimal value $0$ in (\ref{SCmodelI}). This yields a threshold parameter $q_c$ given by
\begin{equation}\label{qcmodelI}
 q_c=\frac{(1-\gamma)P_{no-return}}{\gamma+(1-\gamma)P_{no-return}},
\end{equation}
where 
\begin{equation}\label{noreturn}
P_{no-return}=\left(\frac{1}{(2\pi)^D}\int_{\cal B}\frac{d\vec k}{1-\widehat{p}(\vec k)}\right)^{-1}.
\end{equation}
This latter quantity is recognized as the classic probability that a Markovian random walk with distribution $p(\ell)$ never returns to its starting site on a infinite $D$-dimensional lattice \cite{feller1}.

 \subsubsection{Order parameter} We deduce from above that, if $0\le q\le q_c$, then $P_0=0$ is the only acceptable steady state solution. This corresponds to the delocalized phase, where the particle diffuses with an unbounded MSD [even though  it is logarithmic, see Eqs. (\ref{m2})-(\ref{slowgauss})], so that the density vanishes asymptotically. Above threshold, on the other hand, new localized solutions with $P_n>0$ are possible. For $q$ larger but close to $q_c$, an expansion of Eq. (\ref{SCmodelI}) gives \cite{falcon2017localization}
\begin{equation}
    P_0\sim (q-q_c)^{\beta}.
\end{equation}
Therefore, $P_0$ is analogous to the order parameter of a second order phase transition and the exponent $\beta$ depends on the dimension and on the type of distribution $p({\ell})$ in the random walk mode. If $p({\ell})$ has a second moment, or  $\sum_{{\ell}} |{\ell}|^2p({\ell})<\infty$, like in the nearest neighbour random walk, one obtains
\begin{equation}\label{betaRW}
\beta= \left\{
\begin{array}{cl}
1 &\,\, {\rm for} \quad D\ge4 \\
\frac{2}{D-2} &\,\, {\rm for} \quad 2<D<4 \\
\frac{D}{2-D}&\,\, {\rm for} \quad  D<2.
\end{array}
\right.
\end{equation}
The particular case $D=2$ will be analysed in Section \ref{sec:lowcritd}. If the random displacements are L\'evy flights, or $p({\ell})\sim 1/|{\ell}|^{1+\mu}$ at large $|{\ell}|$ with $0<\mu<2$, then
\begin{equation}\label{betaLevy}
\beta= \left\{
\begin{array}{cl}
1 &\,\, {\rm for} \quad D\ge2\mu  \\
\frac{\mu}{D-\mu} &\,\, {\rm for} \quad \mu<D<2\mu  \\
\frac{D}{\mu-D}&\,\, {\rm for} \quad  D<\mu.
\end{array}
\right.
\end{equation}

\subsubsection{Localization/delocalization transition.} It is quite remarkable that a key property of our memory walk in an inhomogeneous medium, namely, the existence of a phase transition at a critical memory rate $q_c$, depends on the recurrence property of the standard memory-less lattice random walk. From Eq. (\ref{qcmodelI}), $q_c=0$ iff $P_{no-return}=0$. Namely, localized states exist for all $q>0$ if the underlying Markov process in the absence of resetting ($q=0$) is recurrent. This happens when the integral in (\ref{noreturn}) is infinite, due to a divergence at small $|\vec k|$. For random steps of finite variance, $\widehat{p}(\vec k)$ is of the form 
$\widehat{p}(\vec k)\simeq 1-K_2|\vec k|^2$, which implies that $q_c=0$ for $D\le 2$. In other words, the lower critical dimension, denoted as $D_c$, is $2$. Conversely, a phase transition at finite $q_c$ exists only for $D>D_c$. In the case of L\'evy flights,  $\widehat{p}(\vec k)\simeq 1-K_{\mu}|\vec k|^{\mu}$ and the lower critical dimension is $D_c=\mu$. Therefore, the two phases can be observed in $1D$ if $\mu<1$ \cite{falcon2017localization}.

\subsubsection{Localization length exponent.} The behaviour of the localized profile $P_n$ at large $|n|$ can be deduced from the study of $\widehat{P}_k$ in the small $|\vec k|$ limit. For random steps of finite variance, expression (\ref{transPn}) becomes
\begin{equation}\label{ness}
 \widehat{P}_k\simeq\frac{q^*}{K|\vec k|^2+q^*},
\end{equation}
with 
\begin{equation}\label{q*}
 q^*=q\gamma P_0,
\end{equation}
and $K=(1-q)K_2$ a re-scaled diffusion constant. Expression (\ref{ness}) can be expressed as $\widehat{P}_k\sim 1/(|\vec k|^2+\xi^{-2})$, whose inverse transform is proportional to $\exp(-|n|/\xi)$ in all $D$ \cite{evans2014diffusion}. Therefore, the localized profiles are exponential and $\xi$ represents the characteristic extent of the NESS in space. Note that this form in Fourier space is the same as the expression of the correlation function of the Ising model in the Gaussian approximation, where in that context $\xi$ is the correlation length \cite{goldenfeld2018lectures}. Here, one naturally identifies $\xi$ with the localization length, which is deduced from Eqs. (\ref{ness})-(\ref{q*}):
\begin{equation}\label{ximodelI}
\xi=\left(\frac{K}{q^*}\right)^{1/2}=\left(\frac{K}{q\gamma P_0}\right)^{1/2}.
\end{equation}
Owing to the fact that $P_0$ tends to $0$ as $q\rightarrow q_c$ from above, the localization length diverges algebraically, as in a second order phase transition:
\begin{equation}
\xi\sim(q-q_c)^{-\nu},\quad{\rm with}\,\, \nu=\beta/2.
\end{equation}
From (\ref{betaRW}), the correlation length exponent is given by
\begin{equation}\label{nu}
\nu= \left\{
\begin{array}{cl}
1/2 &\,\, {\rm for} \quad D\ge 4  \\
\frac{1}{D-2} &\,\, {\rm for} \quad 2<D<4  \\
\frac{D}{4-2D}&\,\, {\rm for} \quad  D<2.
\end{array}
\right.
\end{equation}
Hence, $D=4$ is the upper critical dimension of this model.

Surprisingly, these results on the localization transition in Model I closely match the well-known phenomenology of the Anderson transition for waves in quenched disordered media \cite{anderson1958absence}. In that problem, apparently unrelated to the present one, electronic or classical waves undergo a transition between diffusive and localized behaviours due to strong interference effects when the disorder strength $\eta$ crosses a critical value $\eta_c$. In $1D$ and $2D$, waves are localized at any $\eta>0$, whereas a finite mobility edge $\eta_c$ separates diffusive and localized states in $3D$.
The expressions (\ref{nu}) for the exponent $\nu$ are actually the same as the ones predicted by the self-consistent theory (SCT) of Anderson localization developed some time ago by Vollhardt and Wolfle \cite{vollhardt1980diagrammatic,vollhardt1982scaling}. Therefore our model (or at least the results derived with the de-correlation approximation) fall in the same universality class as the SCT of Anderson localization, which is also an approximation to the full wave propagation problem. 

\subsubsection{Effective resetting to the impurity and learning.} Expression (\ref{ness}) for the NESS in the localized phase has a simple yet important interpretation. It is identical to the steady state solution of the diffusion equation with diffusion constant $K$ and resetting rate $q^*$ to the origin \cite{evans2011diffusion,kusmierz2015optimal}. In the continuous time and space limit, such equation, first presented in \cite{evans2011diffusion}, reads
\begin{equation}\label{diffreset}
\frac{\partial P(x,t)}{\partial t}=K\Delta P(x,t)-q^*P(x,t)
+q^*\delta(x).
\end{equation}
Equation (\ref{diffreset}) is local in time and markedly different from our original equation (\ref{mastermodelI}).
Therefore, at large times, it is as if the walker of Model I reset to the impurity site {\it only}, instead of using its memory to revisit any previous site. This effective resetting nevertheless occurs at a rate lower than the actual resetting parameter, since $q^*<q$. Eventually, the effective resetting rate $q^*$ vanishes (like $P_0$) at the critical point $q=q_c$. 

The localized behaviour is thus a manifestation of spatial learning: independently of its initial position, the walker ends up by effectively resetting to the resource site, considered as a valuable site.  This resetting site is not set before hand but emerges from the dynamics and the experience of the walker, contrary to Eq. (\ref{diffreset}) where it is explicitly incorporated through the $\delta$-function term. Biologically, localization represents a successful adaptation to the environment. Below $q_c$, the NESS no longer exists: the walker does not use its memory often enough to be able to steadily revisit the impurity and the learning process of its location cannot be completed. The selection at large times of the best option among a set of possibilities is the typical outcome of models of reinforcement learning \cite{luce2012individual}. Our model can be considered as an extension of such models for a searcher moving from site to site in a spatially explicit environment.

\subsection{Lower critical dimension}\label{sec:lowcritd}
We now present new results and start with the analysis of the critical behaviour of the localization transition at the lower critical dimension $D_c$.
\subsubsection{Simple random walks.}
For random walks with finite variance, $D_c=2$, which is a direct consequence of the recurrence/transience transition of these processes. The $2D$ case thus deserves a special attention, apart from its relevance for applications in ecology. In $D=2$, according to Eq. (\ref{nu}), the correlation length exponent $\nu$ becomes infinite, which indicates that the divergence of $\xi$ near $q_c=0$ does not follow a power-law. To obtain the correct behaviour, one can solve the self-consistent equation (\ref{SCmodelI}) for $P_0$ when $q$ is close to $0$, by noting that most of the contribution to the integral comes from the small wave-number region. For the nearest neighbour random walk on the square lattice, $\widehat{p}(\vec k)\simeq 1-|\vec k|^2/4$ at small $|\vec k|$. Changing to polar coordinates and noting $\rho\equiv|\vec k|$, Eq. (\ref{SCmodelI}) becomes
\begin{equation}\label{sct2D}
    \int_0^{R}\frac{d\rho}{2\pi}\frac{\rho}{\frac{1-q}{4}\rho^2+q\gamma P_0}
    \simeq\frac{1-\gamma}{q\gamma(1-\gamma P_0)},
\end{equation}
for $(q,P_0)$ small and where $R$ is an unimportant constant of order $1$. A straightforward integration of (\ref{sct2D}) gives
\begin{equation}
    \ln(q\gamma P_0)\simeq -\frac{\pi(1-\gamma)(1-q)}{q\gamma(1-\gamma P_0)}
    \simeq -\frac{\pi(1-\gamma)}{q\gamma}.
\end{equation}
Therefore, as $q\rightarrow0$, $P_0$ tends to $0$ with an essential singularity:
\begin{equation}\label{P02D}
P_0(q)\sim \frac{1}{q\gamma}\exp\left[-\frac{\pi(1-\gamma)}{\gamma q} \right].
\end{equation}
The localization length stems from relation (\ref{ximodelI}), with the diffusion constant given by $K=(1-q)/4$ here. One deduces
\begin{equation}\label{xi2D}
    \xi\sim\exp\left[ \frac{\pi(1-\gamma)}{2\gamma q}\right].
\end{equation}

\begin{figure}
\begin{center}
\includegraphics*[width=0.6\textwidth]{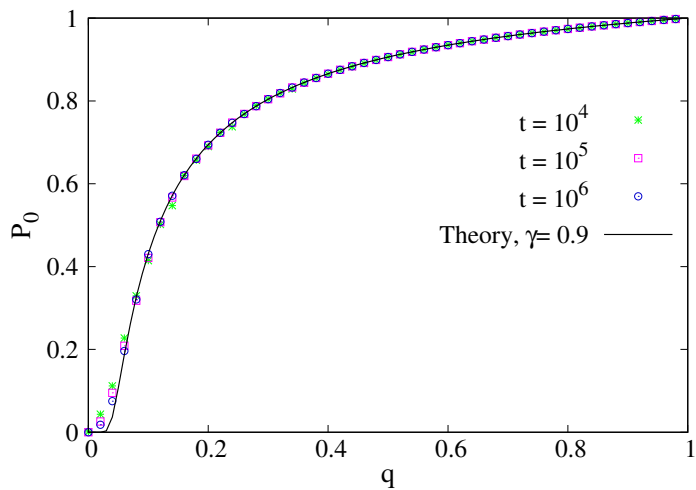}
    \caption{$P_0$ as a function of $q$ for $\gamma=0.9$ and nearest neighbour random steps on a $2D$ square lattice. Symbols are simulation results for walks starting at the origin and performing $t$ steps, and the full line is obtained from Eq. (\ref{SCmodelI}).}
    \label{fig2Dgamma}
\end{center}
\end{figure}

Hence, $\xi$ diverges faster than any power-law in the low memory limit. The scaling theory of Anderson localization in two dimensions predicts a similar law for the correlation length, as a function of the characteristic dimensionless conductivity parameter \cite{abrahams1979scaling}. 
In Fig. \ref{fig2Dgamma} we compare the numerical solution of the full self-consistent equation (\ref{SCmodelI}) with Monte Carlo simulations, obtaining a very good agreement. This justifies a posteriori the de-correlation approximation used to pass from Eqs. (\ref{mastermodelI}) to (\ref{nessmodelI}). What seems to be an abrupt transition at a small $q_c$ in the theoretical curve actually corresponds to a smooth behaviour where $P_0$ is very small but not zero, typical of an essential singularity. Near $q=0$, the convergence of the simulations towards the stationary profile is very slow: we interpret this by the fact that $\xi$ is very large in this limit, therefore the walks must diffuse for long distances before reaching the tails of the NESS. This process takes a long time because of the logarithmic diffusive dynamics that characterizes the model without impurity.

\subsubsection{L\'evy flights.}\label{qcLF}

We now consider heavy-tailed step distributions between relocation events, of the form $p({\ell})\sim 1/|{\ell}|^{1+\mu}$ at large $|{\ell}|$, with $0<\mu<2$. The Fourier transform of $p({\ell})$ is given by $\widehat{p}(\vec k)=1-K_{\mu}|\vec k|^{\mu}$ at small $|\vec k|$, with $K_{\mu}$ a constant. We recall that, from Eqs. (\ref{qcmodelI})-(\ref{noreturn}),  $q_c=0$ if $D\le\mu$ and $q_c>0$ if $D>\mu$. Therefore, the critical dimension is now
\begin{equation}
    D_c=\mu.
\end{equation}
This is the dimension at which the recurrence/transience transition occurs for L\'evy flights with index $\mu$. Right at $D=D_c$ and for small $q$, Eq. (\ref{SCmodelI}) now reads
\begin{equation}
    \frac{S_{\mu}}{(2\pi)^{\mu}} \int_0^{R}d\rho\frac{\rho^{\mu-1}}{(1-q)K_{\mu}\rho^{\mu}+q\gamma P_0}
    \simeq\frac{1-\gamma}{q\gamma(1-\gamma P_0)},
\end{equation}
with $S_{\mu}=\mu\pi^{\frac{\mu}{2}}/\Gamma(\frac{\mu}{2}+1)$ the area of the sphere of unit radius in dimension $\mu$. After integration one obtains
\begin{equation}
    P_0(q)\sim \frac{1}{q\gamma}
    \exp\left[-2^{\mu}\pi^{\frac{\mu}{2}}\Gamma(\frac{\mu}{2}+1)K_{\mu}\frac{1-\gamma}{\gamma q} \right],
\end{equation}
at small $q$. This relation generalizes Eq. (\ref{P02D}), which is recovered for $\mu=2$ and $K_{\mu}=1/4$. Although $D_c$ is not an integer dimension in general, the behaviour of $P_0$ with $q$ at $D_c$ is still governed by an essential singularity.

\begin{figure}
\begin{center}
\includegraphics*[width=0.47\textwidth]{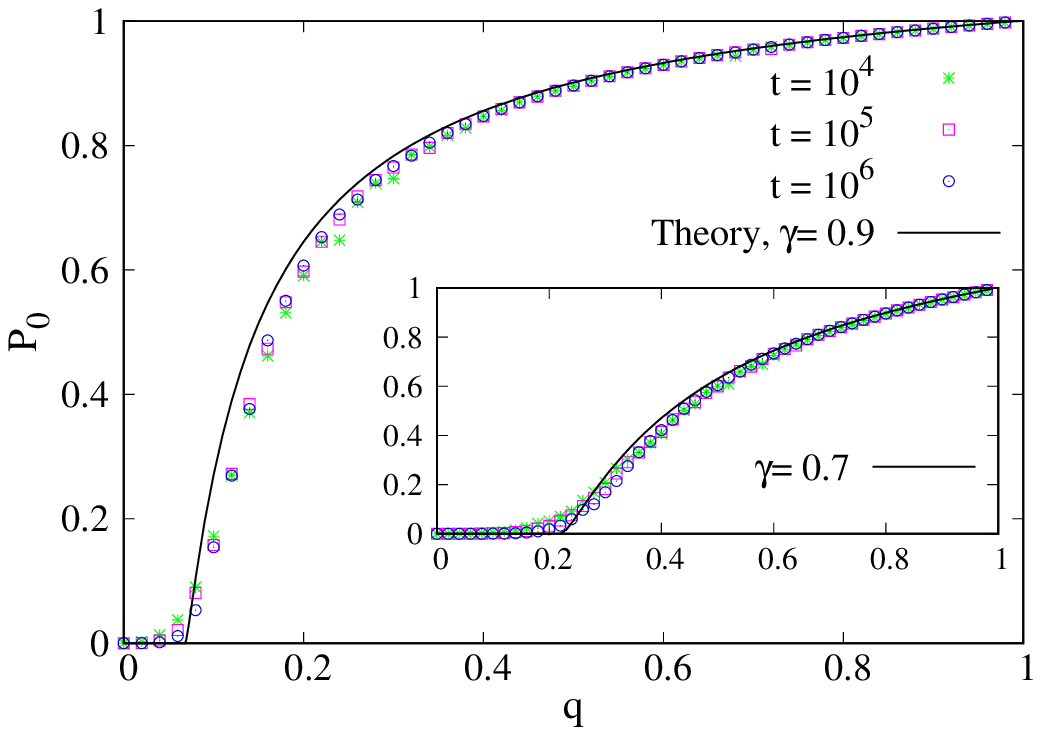}
\includegraphics*[width=0.47\textwidth]{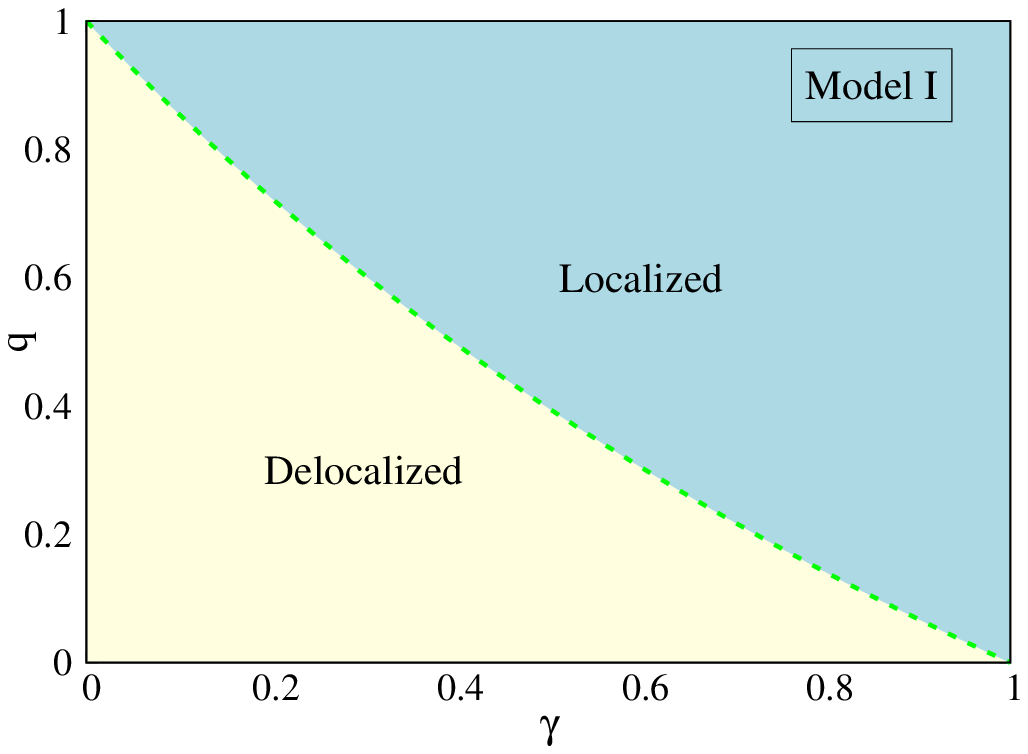}
    \caption{{\bf Left:} Same as Figure \ref{fig2Dgamma} for a $3D$ cubic lattice. Inset: $\gamma=0.7$. {\bf Right:} Phase diagram of Model I in $3D$ for nearest neighbour random steps.}
    \label{fig3Dgamma}
\end{center}
\end{figure}

\subsection{$3D$ case}

$D=3$ is the physical dimension where a phase transition at finite $q_c$ can be observed in our model with a nearest neighbour random walk. On the cubic lattice, Eq. (\ref{SCmodelI}) and Monte Carlo simulations once again exhibit a good agreement for all values of $q$, as shown in Fig. \ref{fig3Dgamma}-Left. From relation (\ref{qcmodelI}) and the value $P_{no\ return}=0.6595...$, one obtain $q_c=0.0682...$ for $\gamma=0.9$ and $q_c=0.2203...$ for $\gamma=0.7$ (inset). Although deviations from theory are most noticeable near $q_c$ (like in $2D$), there is a clear numerical evidence of a phase transition. 
Fig. \ref{fig3Dgamma}-Right displays the phase diagram of Model I in $3D$, drawn from Eq. (\ref{qcmodelI}).

\section{Results on Model II} 

\subsection{Main equations}

According to the rules enunciated in Section \ref{secmodel} for Model II, the master equation is now given by
    \begin{eqnarray}\label{mastermodelII}
     P_n(t+1) &=& \sum\limits_{{\ell}} (1-r_{n-{\ell}})P_{n-{\ell}}(t)p({\ell})\nonumber \\
              &+& \frac{q}{t+1}\sum\limits_{t'=0}^{t} \mbox{Prob}[X_{t'}=n\;\;\mbox{and}\;\;X_t\neq 0]\nonumber\\
              &+& \frac{\delta q}{t+1}\sum\limits_{t'=0}^{t} \mbox{Prob}[X_{t'}=n\;\;\mbox{and}\;\;X_t= 0]
    \end{eqnarray}
in any dimension $D$. Recall that $P_n(t)$ is the probability of occupying the lattice site $n$ at time $t$ and $p({\ell})$ the probability of performing a displacement ${\ell}$ in the random walk mode. The position dependent resetting probability is given by $r_n=q$ for $n\neq0$ and $r_n=\delta q$ for $n=0$. As for Model I, we use the de-correlation approximation in order to derive analytical results. We approximate $\mbox{Prob}[X_{t'}=n\;\;\mbox{and}\;\;X_t\neq 0]$ by $P_n(t')[1-P_0(t)]$, and $\mbox{Prob}[X_{t'}=n\;\;\mbox{and}\;\;X_t= 0]$ by $P_n(t')P_0(t)$.  The validity of such approximation will be checked with Monte Carlo simulations. Taking jointly the limit $t\rightarrow\infty$ and $t'\rightarrow\infty$, and setting $\lim_{t\rightarrow\infty}P_n(t)=P_n$, we obtain an equation that describes the NESS:
    \begin{equation}
     P_n = (1-q)\sum\limits_{{\ell}}P_{n-{\ell}}p({\ell}) + P_0p(n)q(1-\delta) + qP_n[1-P_0(1-\delta)].
    \end{equation}
Applying the discrete Fourier transform to this equation, one obtains    
\begin{equation}\label{pkmodelII}
     \widehat{P}_k = \frac{P_0q(1-\delta)\widehat{p}(\vec k)}{(1-q)[1-\widehat{p}(\vec k)]+q(1-\delta)P_0}.
\end{equation}
Besides the trivial solution $P_0=0$, other solutions are obtained by substituting (\ref{pkmodelII}) into the general self-consistent relation (\ref{SCgen}). Solutions with $P_0>0$ obey the equation   
    \begin{equation}\label{scP0modelII}
     \frac{1}{(2\pi)^D}\int_{\cal B} \frac{d\vec k}{(1-q)[1-\widehat{p}(\vec k)]+q(1-\delta)P_0} = \frac{1-\delta q}{q(1-\delta)[1-q+q(1-\delta)P_0]}.
    \end{equation}
Once again, fixing $\delta$, the critical point $q_c$ that characterizes the onset of localized behaviour is obtained by setting $P_0=0$ above. After re-arranging terms, one obtains   
    \begin{equation}\label{qcmodelII}
     q_c = \frac{P_{no-return}}{(1-\delta)+\delta P_{no-return}}.
    \end{equation}
Like in Model I, the existence of a phase transition at finite $q_c>0$ is possible when the underlying Markov process (for $q=0$) is transient, that is $P_{no-return}\neq 0$, whereas $q_c=0$ for recurrent processes, that is $P_{no-return}=0$. Eqs. (\ref{pkmodelII}) and (\ref{scP0modelII}) have the same structure of Eqs. (\ref{transPn}) and (\ref{SCmodelI}) for Model I, therefore the localisation transition belongs to the same class. In particular, the large scale behaviour of the NESS obeys (\ref{ness}) with $q^*=q(1-\delta)P_0$ and the divergence of the localization length is given by the scaling laws (\ref{nu}).

\subsection{Analytic expression in $1D$}
As for Model I \cite{falcon2017localization}, it is possible to derive a close solution for the $1D$ nearest neighbour random walk in Model II. We consider the random step distribution
\begin{equation}
     p({\ell}) = \frac{1}{2}\left[\delta_{{\ell},1} + \delta_{{\ell},-1}\right],
\end{equation}
whose Fourier transform is $\widehat{p}(k)=\cos k$. The stationary state (\ref{pkmodelII}) reads
\begin{equation}\label{pkII1D}
	\widehat{P}_k = \frac{P_0q(1-\delta)\cos k}{(1-q)(1-\cos k)+q(1-\delta)P_0}
	             = A + \frac{B}{(1-q)(1-\cos k)+q(1-\delta)P_0}
\end{equation}
with $A = [q(1-\delta)P_0]/(q-1)$ and $B = A\;[q-1-q(1-\delta)P_0]$.
The inverse Fourier transform of this expression can be performed thanks to the identity
\begin{equation}\label{coskint}
	\frac{1}{2\pi}\int\limits_{-\pi}^{\pi} dk \frac{\cos (kn)}{1+a^2-2a\cos k} = \frac{1}{(a^2-1)a^{|n|}}
\end{equation}
for $a^2>1$. We hence write the denominator in Eq. (\ref{pkII1D}) under the form $b(1+a^2-2a\cos k)$. By identification,
\begin{eqnarray}
	&&2ab=1-q\\
	&&b(1+a^2)=1-q[1+(1-\delta)P_0],
\end{eqnarray}
which gives
\begin{equation}\label{a}
	a = 1 + u + \sqrt{u(2+u)}
\end{equation}
with $u = [q(1-\delta) P_0]/(1-q)$.
Using Eqs. (\ref{pkII1D})-(\ref{coskint}), $P_n$ is given by
\begin{equation}
	\frac{1}{2\pi}\int\limits_{-\pi}^{\pi}dk \left(A+\frac{B}{b(1+a^2-2a\cos k)}\right)\cos (kn) = A \delta_{n,0}+ \frac{2aB}{(1-q)(a^2-1)a^{|n|}},
\end{equation} 
or,
\begin{equation}\label{pnII1D}
	P_n =  \frac{q(1-\delta)P_0}{q-1}\delta_{n,0} + \frac{q(1-\delta)P_0[1-q+q(1-\delta)P_0]}{(1-q)^2}\frac{2a}{(a^2-1)a^{|n|}}.
\end{equation}
(Again, the Kroneker symbol $\delta_{n,0}$ must not be confused with the parameter $\delta$). Setting $n=0$ in Eq. (\ref{pnII1D}), a self-consistent equation is obtained for the unknown probability $P_0$ of occupying the origin,
\begin{equation}
	2q(1-\delta)[q-1-q(1-\delta)P_0] = (\delta q - 1)(1-q)(a-a^{-1}).
\end{equation} 
Inserting expression (\ref{a}) for $a$ into the above equality yields a quadratic equation for $P_0$:
\begin{equation}\label{quadP0}
	a_1 P_0^2 + b_1 P_0 + c_1 = 0
\end{equation}
with,
\begin{eqnarray}
	a_1 &=& q^2(1-\delta)^2(1-q)[1+q(1-2\delta)]\\
	b_1 &=& 2q(1-\delta)(1-q)^2[1+q(1-2\delta)]\\
	c_1 &=& -q^2(1-\delta)^2(1-q)^2.
\end{eqnarray}
Equation (\ref{quadP0}) has only one positive root given by
\begin{equation}\label{solP01D}
	P_0 = \frac{1}{q(1-\delta)}\left[
	\sqrt{\frac{(1-q)(1-\delta q)^2}{1+q(1-2\delta)}}+q-1 \right].
\end{equation}

We comment on several properties of the above expression. First, it is easy to show that $0\le P_0\le 1$ for any $(q,\delta)\in [0,1]^2$. Fixing $\delta$,
an expansion of (\ref{solP01D}) at small $q$ yields
\begin{equation}
P_0\simeq \frac{1-\delta}{2}q\rightarrow 0, 
\end{equation}
indicating that the NESS disappear at $q=0$. In this limit, the model reduces to the simple nearest neighbour random walk, which, as is well known, occupies the origin with probability zero at $t=\infty$. Conversely, when $q\rightarrow 1$ it is easy to see from (\ref{solP01D}) that 
\begin{equation}\label{P0q1}
P_0\simeq\sqrt{\frac{1-q}{2(1-\delta)}}
\end{equation}
for any fixed $\delta<1$. Therefore $P_0\rightarrow0$ as well in the strong memory limit, which may sound surprising. Indeed, the walker that starts at the origin will sooner or later jump to a  nearest neighbour site ($n=1$ or $-1$) that will be reinforced after many jumps, leaving the origin unoccupied. By evaluating Eq. (\ref{pnII1D}) at $n=\pm 1$ and using (\ref{P0q1}), one actually obtains
\begin{equation}
P_1=P_{-1}=\frac{1}{2},    
\end{equation}    
\begin{figure}
\begin{center}
\includegraphics*[width=0.5\textwidth]{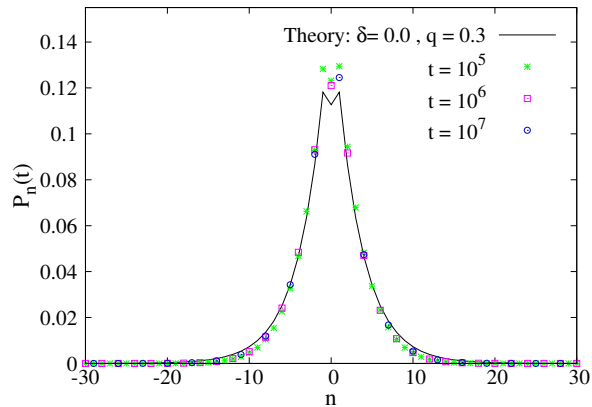}
    \caption{Localized profile $P_n$ as a function of the position $n$ in $1D$ for Model II, as given by Eq.  (\ref{pnII1D}). Symbols are Monte Carlo simulation results for walks starting at the origin and performing $t$ steps.}
    \label{figPnmodelII}
\end{center}
\end{figure}
in the limit $q=1$, whereas $P_n=0$ for $|n|>1$ due to the divergence of the constant $a$.  Therefore, the particle is fully localized on the sites $n=1$ and $n=-1$ when $q\rightarrow 1$, and not at the origin like in Model I. If the starting site in not the origin, the particle remains there forever, as a trivial consequence of the dynamics. The self-consistent theory presented here cannot be valid for trajectories that do not have the opportunity to visit the impurity. Figure \ref{figPnmodelII} displays the NESS obtained for $q=0.3$ and $\delta=0$, and where it is already clear that $P_{n=0}<P_{n=1}$ asymptotically. An excellent agreement over the whole range of position $n$ is obtained between theory and Monte Carlo simulations of the memory walks.

\begin{figure}
\begin{center}
\includegraphics*[width=0.45\textwidth]{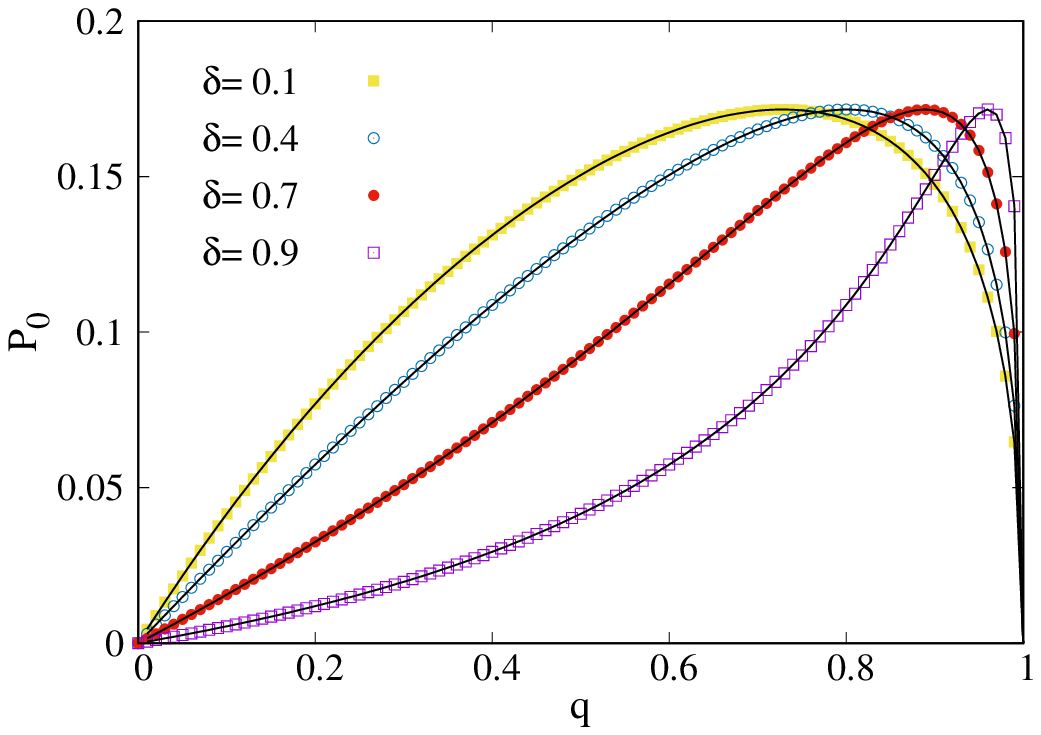}
\includegraphics*[width=0.45\textwidth]{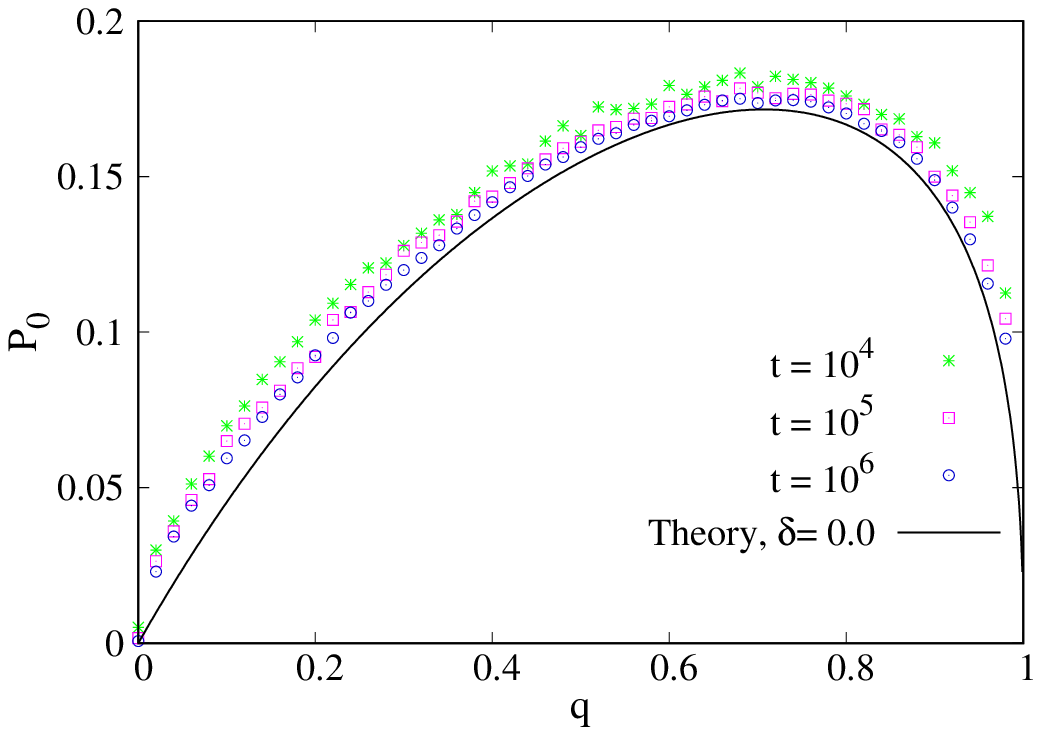}
    \caption{{\bf Left:} $P_0$ as a function of $q$ in $1D$ for several values of $\delta$ as given by Eq. (\ref{solP01D}). {\bf Right:} Same quantity for $\delta=0$ (solid line); the symbols are simulations.}
    \label{fig1D}
\end{center}
\end{figure}

One deduces from the above considerations that, contrary to Model I, $P_0$ is non-monotonic with $q$ and must have a maximum for a particular value $q_{\delta}^*$ of the resetting probability. Figure \ref{fig1D}-Left  displays the solution (\ref{solP01D}) as a function of $q$ for several $\delta$. The location of the maximum, $q_{\delta}^*$, increases with $\delta$: overall, localization is weakened at larger $\delta$.

Of particular interest is the case $\delta = 0$: in this situation the walker does not use its memory at all when it occupies the origin (although it remembers this position for future relocations) and performs random jumps to the sites $n=1$ or $n=-1$ instead. Equation (\ref{solP01D}) with $\delta=0$ reduces to
\begin{equation}\label{solP01Dgamma0}
    	P_0 = \frac{1}{q}\left[
	\sqrt{\frac{1-q}{1+q}}+q-1 \right],
\end{equation}
which is maximal at $q=q_0^*=1/\sqrt{2}=0.707106...$. At its maximum, $P_0=(\sqrt{2}-1)^2=0.171572...$, and this value is also the maximum reached by $P_0$ for any parameter $\delta\neq0$ (see Fig. \ref{fig1D}-Left). This probability is  significantly smaller than the values close to unity that can be attained in Model I. A quantitative agreement is obtained between Monte Carlo simulations and
 Eq. (\ref{solP01Dgamma0}), see Fig. \ref{fig1D}-Right, suggesting that the de-correlation approximation on which all these analytic results rely might be exact. As the simulation time $t$ increases, the numerical results slowly approach the expected asymptotic curve. The slow convergence is attributed to the logarithmic dynamics that govern the preferential visit model, as previously discussed. 

We finally comment on the case where $q$ is fixed and $\delta$ close to unity. An expansion of Eq. (\ref{solP01D}) with $\delta=1-\epsilon$ and $\epsilon\ll 1$ gives $P_0\propto \epsilon$. Hence, when the difference between the resetting rates at the origin and at the other sites tends to zero, the NESS disappears. This is in agreement with the behaviour found in the original model without impurity, where the mean square displacement grows unbounded as $\ln t$ for any value of $q$, implying a vanishing density at $t=\infty$.

\subsection{Analysis of the $2D$ and $3D$ cases}

\begin{figure}
\begin{center}
\includegraphics*[width=0.45\textwidth]{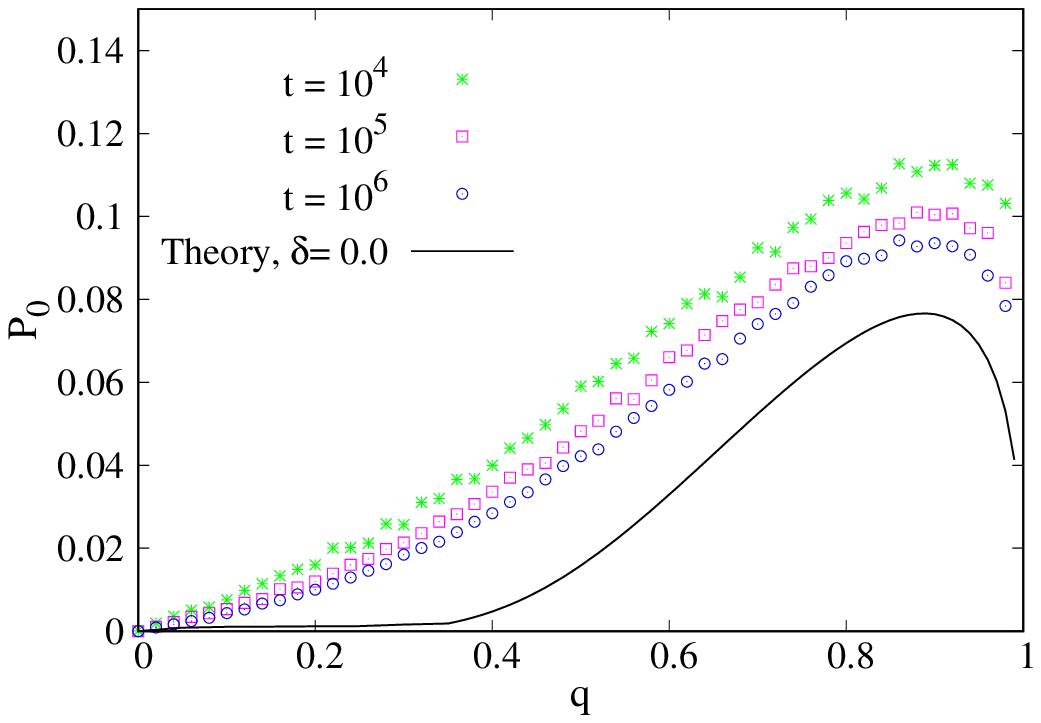}
\includegraphics*[width=0.45\textwidth]{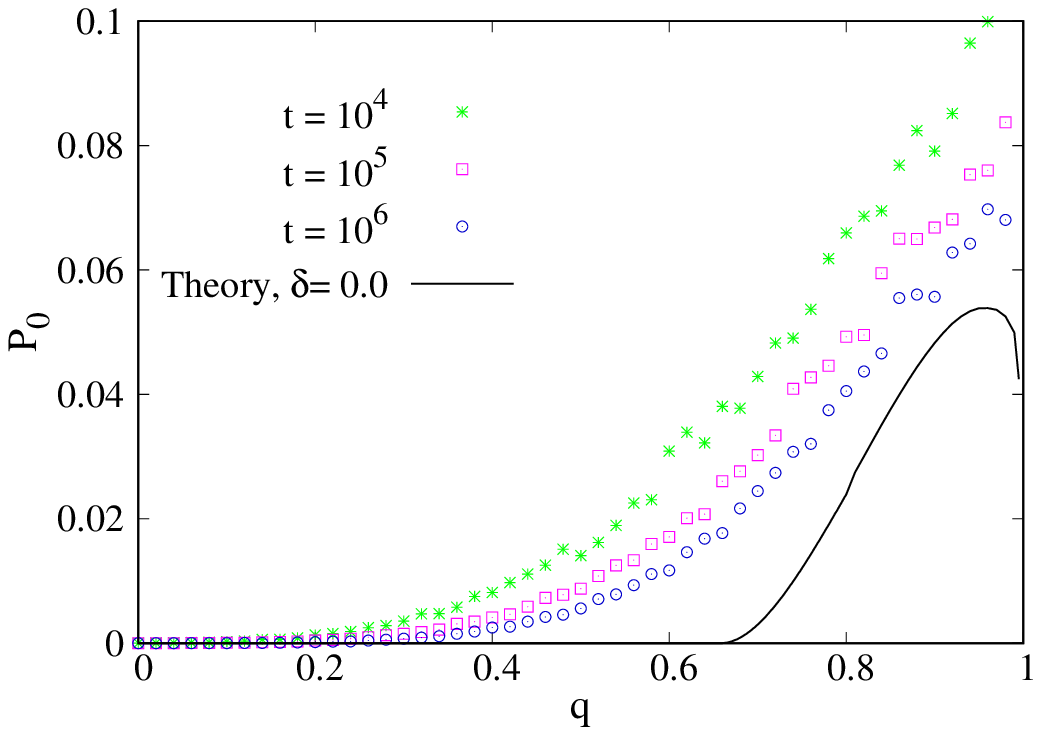}
\end{center}
\hspace{4.5cm}\includegraphics*[width=0.45\textwidth]{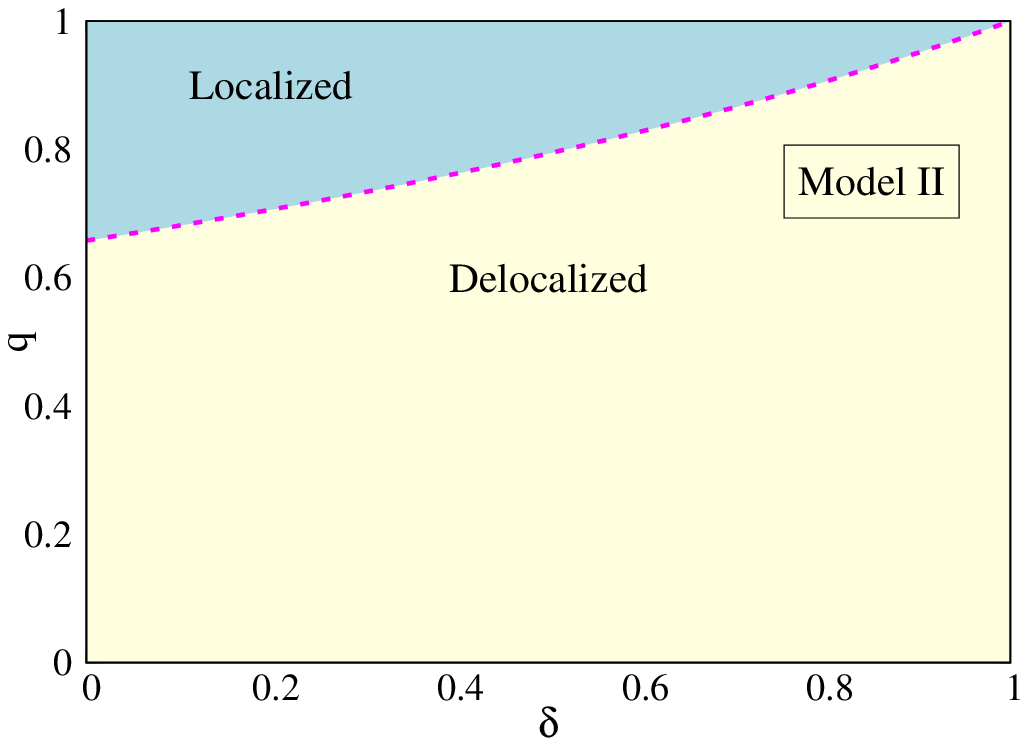}
\begin{center}
    \caption{{\bf Left:} $P_0$ as a function of $q$ for $\delta=0$ in Model II with  nearest neighbour random steps on the $2D$ square lattice. {\bf Right:} $3D$ case. Symbols are Monte Carlo simulations and the solid lines, numerical solution of the self-consistent equation (\ref{scP0modelII}). {\bf Bottom:} $3D$ phase diagram.}
    \label{fig2D3Ddelta}
\end{center}
\end{figure}

We proceed with a study of Model II with nearest neighbour random walks in higher dimensions. $D=2$ corresponds to the lower critical dimension, $q_c$ is still $0$ and the correlation length can be obtained by following the same route leading to Eq. (\ref{xi2D}) for Model I. Starting this time from Eq. (\ref{scP0modelII}) we obtain after simple algebra
\begin{equation}\label{P02DmodelII}
P_0\sim\frac{1}{q(1-\delta)}\exp\left[-\frac{\pi}{q(1-\delta)}\right],
\end{equation}
for $q\ll1$. The small $k$ behaviour of Eq. (\ref{pkmodelII}) allows us to identify the correlation length as $\xi=\{D/[q(1-\delta)P_0]\}^{1/2}$. Therefore, at small resetting rates,
\begin{equation}\label{xi2DmodelII}
\xi\sim \exp\left[\frac{\pi}{2(1-\delta)q}\right].
\end{equation}
Figure \ref{fig2D3Ddelta}-Left displays $P_0$ vs. $q$ for $\delta=0$. The numerical simulations exhibit the same qualitative behaviour as the theory, but the finite time effects are very strong and a quantitative comparison is no longer possible.
One can notice however that the characteristic resetting rate $q_{char}$ which describes the essential singularity is significantly larger in Fig. \ref{fig2D3Ddelta}-Left than in Fig. \ref{fig2Dgamma} for Model I. This feature can be understood from Eq. (\ref{xi2DmodelII}), which predicts that  $q_{char}=\pi/[2(1-\delta)]=\pi/2$ (for $\delta=0$), to be compared, for Model I, with $q_{char}=\pi(1-\gamma)/(2\gamma)=\pi/18$ (for $\gamma=0.9$).

An essential singularity at $q=0$ like in Eq. (\ref{P02DmodelII}) is also observed for L\'evy flights of index $\mu$ at their critical dimension $D_c=\mu$. Following the same steps as in Section \ref{qcLF}, one obtains
\begin{equation}
P_0\sim\frac{1}{q(1-\delta)}
\exp\left[-2^{\mu}\pi^{\frac{\mu}{2}}\Gamma(\frac{\mu}{2}+1)K_{\mu}\frac{\pi}{q(1-\delta)}\right],
\end{equation}
at small $q$.

For nearest neighbour random walks on the $3D$ cubic lattice, $q_c=0.6595...$ from Eq. (\ref{qcmodelII}) with $\delta=0$. The full curve $P_0(q)$ is displayed in Fig. \ref{fig2D3Ddelta}-Right, along with simulation results. Once again the agreement is only qualitative, presumably because of the extremely slow convergence to the NESS. Note also that the occupation probabilities become small ($<0.06$ for all $q$), which makes their numerical estimates more difficult. Figure \ref{fig2D3Ddelta}-Bottom displays the phase diagram of Model II in $3D$ as predicted by Eq. (\ref{qcmodelII}).

\section{Absence of localization in Model III} 

To illustrate the importance of a space dependent resetting rate on the localized states in Models I and II, we now analyze Model III, where $r_n$ is uniform and equal to $q$.  We particularly focus on the behavior of the impurity occupation probability $P_0(t)$ at large time in $1D$, and show that this quantity always vanishes as $t\rightarrow\infty$.

In marked contrast to Model I and II, Model III obeys an exact master equation which involves solely the single-time distribution $P_n(t)$. For nearest neighbour jumps, one has
\begin{eqnarray}
P_n(t+1)&=& \frac{1-q}{2}\,\left[P_{n+1}(t)+P_{n-1}(t)\right]+ 
\frac{q}{t+1}\,\sum_{t'=0}^{t} P_n(t')\nonumber\\
&&-\Delta [\delta_{n,1}+\delta_{n,-1}-2\delta_{n,0}]P_0(t)
\, ,
\label{evol.2}
\end{eqnarray} 
where 
\begin{equation}
\Delta=\frac{\gamma(1-q)}{2}.
\label{Delta}
\end{equation}
 Written this way, the effect of the impurity is incorporated by adding a term to the homogeneous equation \cite{kalay2008effects,kenkre2008molecular}. The last term in the r.h.s. of Eq. (\ref{evol.2}) is considered as a time dependent inhomogeneous function which will be determined self-consistently.
 One can notice the simpler structure of Eq. (\ref{evol.2}), which is linear in $P_n(t)$, when compared to Eqs. (\ref{mastermodelI}) or (\ref{mastermodelII}), that involve multiple-times distribution functions. 
 
 We consider a walker initially located at $n_0$, or $P_n(t=0)=\delta_{n,n_0}$, and define the discrete Laplace transform
 \begin{equation}
 {\tilde P_n}(s)= \sum_{t=0}^{\infty} P_n(t) s^t.
 \end{equation}
 By multiplying Eq. (\ref{evol.2}) by $s^t$ and summing over $t$ from $0$ to $\infty$ gives (see \ref{appa})
 \begin{eqnarray}
{\tilde P_n}(s)&=&\delta_{n,n_0}+\frac{1-q}{2}s
[{\tilde P_{n+1}}(s)+ {\tilde P_{n-1}}(s)]
+q\int_0^{s}du \frac{{\tilde P_n}(u)}{1-u}\nonumber \\
&&-\Delta [\delta_{n,1}+\delta_{n,-1}-2\delta_{n,0}]s{\tilde P_0}(s)
\end{eqnarray}
We define the Fourier-Laplace transform as
\begin{equation}\label{fourierlaplace}
\widehat{\tilde P}(k,s) = \sum_{n=-\infty}^{\infty} e^{-i k n} {\tilde P}_n(s).
\end{equation}
Applying $\sum_{n=-\infty}^{\infty} e^{-i k n}(\cdot)$ to 
Eq. (\ref{evol.2}) yields
\begin{equation}
(1-b_k s) \widehat{\tilde P}(k,s)= e^{-ikn_0} + q \, \int_0^s \frac{\widehat{\tilde P}(k,u)}{1-u}\, du
+2\Delta [1-\cos(k)]s{\tilde P_0}(s),
\label{fk2.3}
\end{equation}
where $b_k=(1-q)\cos(k)$. As shown in the \ref{appa}, the solution of Eq. (\ref{fk2.3})  without impurity ($\Delta=0$) is given by
\begin{equation}\label{Pknoimpur}
\widehat{\tilde P}^{(0)}(k,s) =e^{-ikn_0} (1-s)^{-\alpha_k}\,(1-b_k\,s)^{-(1-\alpha_k)},
\end{equation}
where 
\begin{equation}
\alpha_k= \frac{q}{1-b_k}= \frac{q}{1-(1-q)\,\cos(k)}.
\end{equation}
By taking the derivative of (\ref{fk2.3}) with respect to $s$
and solving the resulting ordinary differential equation for $\widehat{\tilde P}(k,s)$  by the method of variation of constants, one obtains:
\begin{equation}
\widehat{\tilde P}(k,s)=\widehat{\tilde P}^{(0)}(k,s)\left[1+2\Delta (1-\cos k)\int_0^sdu\left(
\frac{1-u}{1-b_ku}\right)^{\alpha_k}\frac{d}{du}
[u{\tilde P_0}(u)]\right].
\label{fk2.4}
\end{equation}
By definition, ${\tilde P_{0}}(s)$ is the inverse Fourier transform of $\widehat{\tilde P}(k,s)$ evaluated at $n=0$,
\begin{equation}
{\tilde P_{0}}(s)=\frac{1}{2\pi}\int_{-\pi}^{\pi}dk\ \widehat{\tilde P}(k,s) ,
\end{equation}
This expression combined with Eq. (\ref{fk2.4}) leads to a self-consistent equation for ${\tilde P_0}(s)$:
\begin{equation}
{\tilde P_0}(s)={\tilde P_0}^{(0)}(s) +\frac{\Delta}{\pi} \int_{-\pi}^{\pi}dk\ \widehat{\tilde P}^{(0)}(k,s)
(1-\cos k)\int_0^sdu\left(
\frac{1-u}{1-b_ku}\right)^{\alpha_k}\frac{d}{du}[u{\tilde P_0}(u)] ,
\end{equation}
where ${\tilde P_0}^{(0)}(s)$ denotes the Laplace transform of $P_{n=0}^{(0)}(t)$ in the impurity-free problem. After an integration by part this expression can be re-written as
\begin{eqnarray}
&&{\tilde P_0}(s)={\tilde P_0}^{(0)}(s) +s{\tilde P_0}(s)\frac{\Delta}{\pi} \int_{-\pi}^{\pi}dk\frac{1-\cos k}{1-b_k s}\nonumber\\
&&-\frac{\Delta}{\pi} \int_{-\pi}^{\pi}dk\ \widehat{\tilde P}^{(0)}(k,s)(1-\cos k)
\int_0^sdu\ u{\tilde P_0}(u) \frac{d}{du}\left(\frac{1-u}{1-b_ku}\right)^{\alpha_k}.
\label{fk2.5}
\end{eqnarray}
We note that, due to the attracting nature of the defect, one must have $P_0(t)\ge P_0^{(0)}(t)\simeq C_0/\sqrt{\ln t}$ with $C_0$ a constant, from Eqs. (\ref{m2})-(\ref{slowgauss}). On the other hand, $P_0(t)$
at most reaches a positive constant $C_1\le 1$ if the walker is localized at $t=\infty$. Thus,
\begin{equation}\label{bound}
\frac{C_0}{(1-s)\sqrt{-\ln(1-s)}}\le {\tilde P_0}(s)\le\frac{C_1}{1-s}
\end{equation}
near $s=1$. To obtain the precise large time behavior of $P_0(t)$, we take the limit $s\rightarrow 1$ in Eq. (\ref{fk2.5}). Since $b_k< 1$ and $0<\alpha_k\le 1$, at leading order
\begin{equation}
\int_0^sdu\ u{\tilde P_0}(u) \frac{d}{du}\left(\frac{1-u}{1-b_ku}\right)^{\alpha_k}\simeq
-\frac{\alpha_k}{(1-b_k)^{\alpha_k}}\int_c^sdu\ {\tilde P_0}(u) (1-u)^{\alpha_k-1}.
\label{fk2.6}
\end{equation}
The integral in (\ref{fk2.6}) diverges as $s\rightarrow1$  because of (\ref{bound}), which is the reason why we have introduced an (unimportant) constant $c<1$. Since ${\tilde P_0}(s)$ is of the form $f(s)(1-s)^{-1}$ with $f(s)$ a slowly varying function, this integral can be expressed as
\begin{equation}
\int_c^sdu\ {\tilde P_0}(u)(1-u)^{\alpha_k-1}\simeq \frac{(1-s)^{\alpha_k}}{1-\alpha_k}{\tilde P_0}(s)\, ,
\end{equation}
\begin{figure}
\begin{center}
\includegraphics*[width=0.45\textwidth]{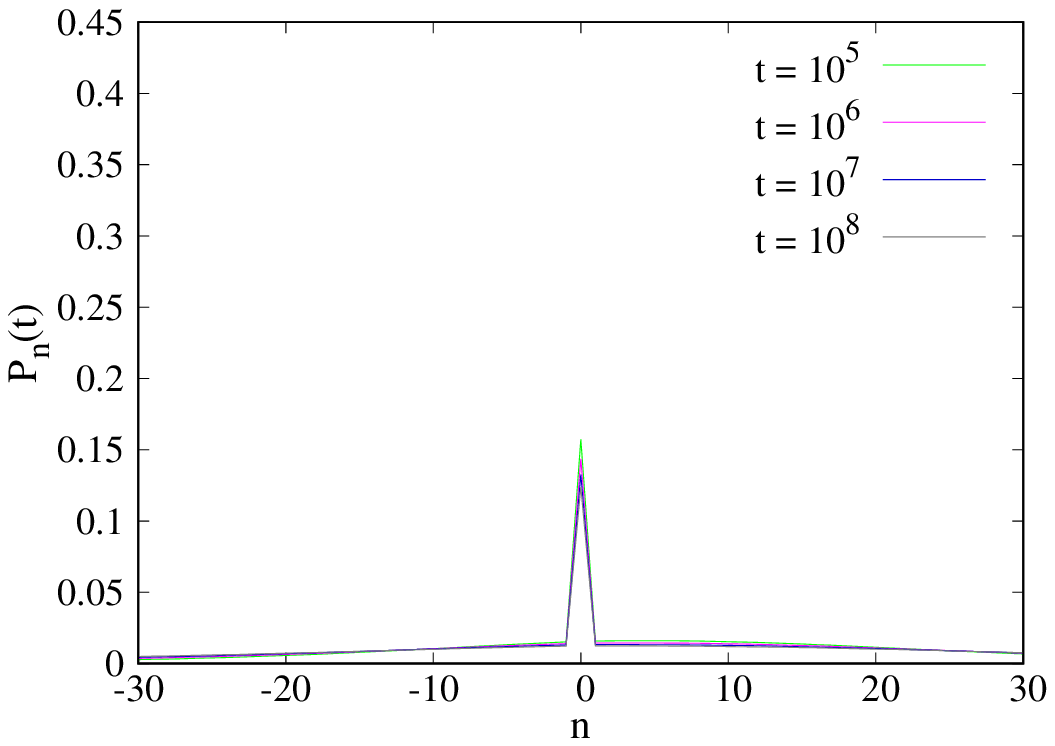}
\includegraphics*[width=0.45\textwidth]{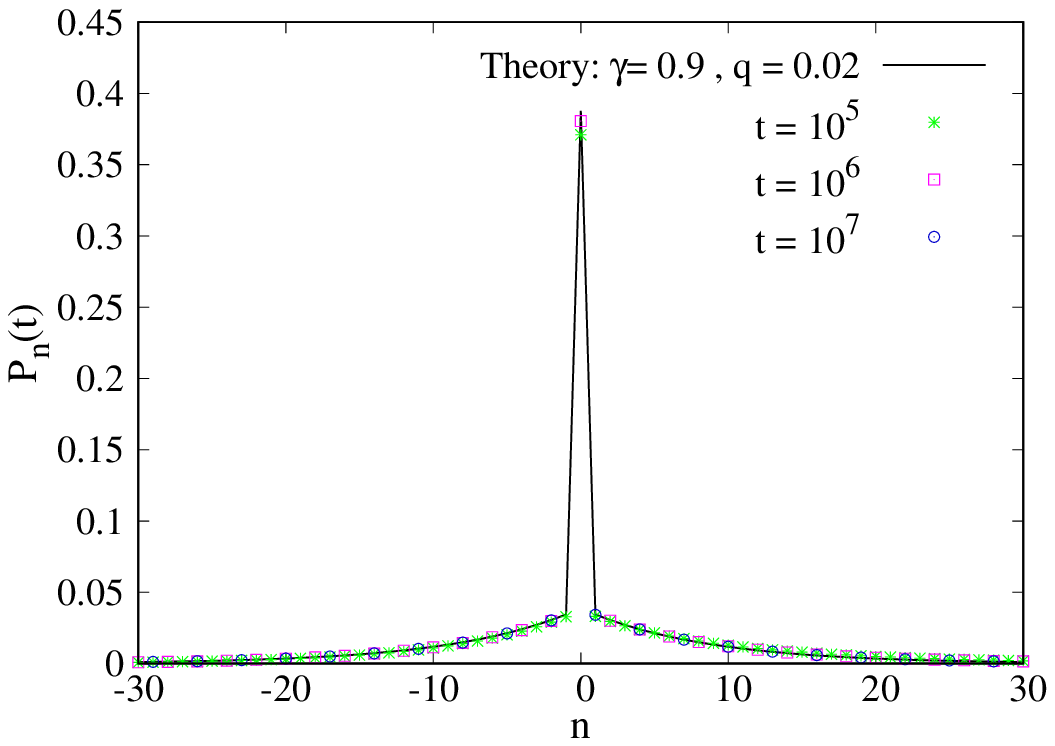}
    \caption{Distribution $P_n(t)$ at different times for a particle with initial position $n_0=5$, $q=0.02$ and $\gamma=0.9$ {\bf Left:} Absence of localization in Model III. {\bf Right:} Localized profile in Model I (dots are simulation results).}
    \label{figPnmodelI}
\end{center}
\end{figure}
Using the exact expression (\ref{Pknoimpur}) for $\widehat{\tilde P}^{(0)}(k,s)$, the last term of Eq. (\ref{fk2.5}) reduces to
\begin{equation}
{\tilde P_0}(s)\frac{\Delta}{\pi}\frac{q}{1-q}\int_{-\pi}^{\pi}dk\ \frac{1}{1-b_k}.
\end{equation}
Equation (\ref{fk2.5}) thus takes the form ${\tilde P_0}(s)\simeq {\tilde P_0}^{(0)}(s)+(c_1+c_2){\tilde P_0}(s)$ near $s=1$. Using (\ref{Delta}) and rearranging terms gives the simple final expression
\begin{equation}
{\tilde P_0}(s)\simeq\frac{{\tilde P_0}^{(0)}(s)}{1-\gamma}.
\end{equation}
Therefore, the presence of the defect amplifies by a factor $1/(1-\gamma)$ the probability of presence at $n=0$, just as in the standard random walk with an impurity. The asymptotic dynamics of $P_{0}(t)$ is thus proportional to that of the defect-free case, for which we know that ${P_0^{(0)}(t)}\sim C_0/\sqrt{\ln t}\rightarrow0$. This implies the absence of localization for any $\gamma<1$ and $q<1$. 

Fig. \ref{figPnmodelI}-Left displays the distribution $P_n(t)$ obtained from solving numerically but exactly the master equation (\ref{evol.2}) for large times. One can notice its (very) slow decay and broadening with time. In contrast, the distribution of Model I with the same parameter values, Fig. \ref{figPnmodelI}-Right, converges towards a narrowly localized NESS. In the latter case, Monte Carlo simulations and the prediction obtained from the numerical inversion of Eq. (\ref{transPn}) are in perfect agreement. This provides further support of the validity of the de-correlation approximation used in the self-consistent theory.

\section{Discussion and conclusions}

We have studied several non-Markovian lattice random walks with preferential resetting to sites visited in the past. The ultra-slow anomalous diffusion that takes place in these systems on homogeneous lattices can be suppressed altogether in the presence of a single impurity site. When the resetting rate $q$ (or rate of memory use) exceeds a critical value, the walker becomes localized around the impurity and the probability distribution of its position adopts a non-equilibrium steady state. The existence of a localized phase depends crucially on the fact that the resetting rate $r_n$ on the impurity is lower than on the other sites, resulting in subtle reinforcement effects around the impurity. Conversely, inhomogeneity in the staying probability $s_n$ seems to play a less important role in the transition.

The processes presented here constitute random walk analogues of the Anderson localization transition. Similarities between the critical properties of reinforced random walks on lattices and the Anderson transition have been drawn in the past \cite{davis1990reinforced,sellke1994reinforced,spencer2010mathematical}, although such comparisons have remained fairly qualitative. Here and in \cite{falcon2017localization}, we have exposed to our knowledge the first precise connection between these two classes of problems, through an analytic study of the phase diagrams, critical points and critical dimensions. The presence of an impurity site, a key ingredient responsible for the phase transition, is usually not considered in the literature on reinforced random walks. 

The genuine Anderson transition is characterized by the absence of diffusion of electron waves in random lattices \cite{anderson1958absence}. In a self-consistent theory (SCT), the quantum diffusion coefficient $K(\omega)$ of the electron density obeys the equation \cite{vollhardt1980diagrammatic,vollhardt1982scaling,wolfle2010self}
\begin{equation}\label{anderson}
    \frac{K(\omega)}{K_0}=1-\eta Dk_F^{2-D}\int_0^{1/\ell}
    dk \frac{k^{D-1}}{-i\omega/K(\omega)+k^2}
\end{equation}
where $\omega$ is the frequency of an a.c. perturbation, $k_F$ the Fermi level, $K_0$ a bare diffusion constant, $\ell$ the mean free path and $\eta$ the standard deviation of the external potential around its mean (disorder strength).  Electrons become localized if $K(\omega)\rightarrow 0$ in the limit $\omega\rightarrow0$, namely if $K(\omega)\simeq -i\omega\times constant$, instead of the usual diffusive behaviour $K(\omega)\simeq K(0)>0$.  One can notice that the self-consistent relation (\ref{anderson}) has the same small $|\vec k|$ structure as our Eqs. (\ref{SCmodelI}) or (\ref{scP0modelII}): the quantity $\lim_{\omega\rightarrow 0}-i\omega/K(\omega)$ is analogous to our $P_0$ and actually represents $\xi^{-2}$, with $\xi$ the localization length of the electrons. Likewise, the disorder strength $\eta$ is analogous to $q$ (or $\gamma$, if $q$ is held fixed in Model I). Consequently,  the solutions of (\ref{anderson}) exhibit the same properties as exposed in Section \ref{sec:previous}, as well as the essential singularity in $D=2$ \cite{wolfle2010self}.

In $3D$, the SCT predicts $\nu=1$ for the correlation length exponent, see Eq. (\ref{nu}), whereas state-of-the-art numerical calculations of electron localization yield the much larger value $\nu=1.571..$, for different choices of the distribution of the random potential \cite{slevin2014critical}. The latter value is also consistent with measurements obtained from experiments in the quantum kicked rotor, a system that exhibits a mapping to the Schr$\ddot{\rm o}$dinger equation with site disorder and which is easier to study \cite{lopez2012experimental}. 

The decoupling approximation used throughout this work and leading to the self-consistent universality class gives results that compare very well with Monte Carlo simulations of the non-Markov processes, in particular in $1D$. Such agreement might be due to the fact that resetting events are non-local in space: a particle can revisit a particular site from any other site, possibly located far away, and not necessarily from a nearest-neighbour site as in standard reinforced random walks \cite{davis1990reinforced}. Therefore the number of visits received by two different sites may exhibit relatively small correlations. A more careful inspection of the critical point through a finite time scaling analysis of the simulations is necessary to validate the SCT exponents. An open question is whether the transition in $3D$ actually belongs to the SCT class, or to the orthogonal universality class with $\nu\simeq 1.57$ mentioned above, or to a third class. It would be also interesting to study standard linearly reinforced random walks with local, nearest neighbour jumps in the presence of a single impurity: the phase diagrams and possible critical properties might differ significantly from processes with stochastic resetting like our Models I and II.

Another outstanding problem is the study of the relaxation dynamics to the stationary state in the localized phase, or the decay at large time of the impurity occupation probability $P_0(t)$ at $q=q_c$. It would be also worth examining how the critical behaviour is  modified if the distribution of intervals between resetting events is no longer exponential, {\it i.e.}, not characterized by a rate $q$. In processes with resetting to the origin, the distribution of times between reset events has an important impact on the non-equilibrium steady states and can optimize mean first passage times \cite{pal2016diffusion,eule2016non,nagar2016diffusion,chechkin2018random}. 

We hope that this study will contribute to the understanding of the mechanisms involved in spatial learning processes by foraging animals. Our results provide a mathematical support to the hypothesis presented some time ago \cite{van2009memory,borger2008there,fagan2013spatial} that memory plays an important role during home range formation. Above a critical resource threshold $\gamma_c$, our model walker builds a NESS, {\it i.e.}, a stationary distribution of space use comparable to a home range in the context of foraging ecology. The diffusive behaviour that takes place below $\gamma_c$ could be advantageous for exploring other regions of space where more valuable resources might be found. Systems with many impurity sites should therefore deserve further study, and could also be used to test the ability of non-Markovian diffusing elements to solve complex optimization problems. Recent numerical results with multiple searchers suggest promising applications in this field \cite{falcon2019collective}.

\ack
DB acknowledges support from DGAPA-PAPIIT Grant IN108318, AFC from a Ph.D CONACYT scholarship, and LG from EPSRC Grant number EP/I013717/1. DB and LG thank the Max Planck Institute for the Physics of Complex Systems for hospitality during a stay where this work was initiated within the Advanced Study Group \lq\lq Anomalous diffusion in foraging\rq\rq.

\appendix

\section{ }\label{appa}

We denote $P_n^{(0)}(t)$ as the probability distribution of the nearest neighbour random walk in $1D$ with preferential resetting, in the absence of impurities. The master equation obeyed by $P_n^{(0)}(t)$ reads
\begin{equation}
P_n^{(0)}(t+1)= \frac{1-q}{2}\,\left[P_{n+1}^{(0)}(t)+P_{n-1}^{(0)}(t)\right]+ 
\frac{q}{t+1}\,\sum_{t'=0}^{t} P_n^{(0)}(t').
\label{appmaster}
\end{equation} 
With the definition of the generating function, ${\tilde P_n}^{(0)}(s)= \sum_{t=0}^{\infty} P_n^{(0)}(t) s^t$, we multiply Eq. (\ref{appmaster}) by $s^t$ and sum over $t$ from $0$ to $\infty$. The left-hand-side gives
\begin{equation}\label{appLHS}
\frac{1}{s}[{\tilde P_n}^{(0)}(s)-\delta_{n,n_0}],
\end{equation}
where we have made use of the initial condition $X_0=n_0$.
The last term of the right-hand-side becomes
\begin{equation}\label{app3rd}
\frac{q}{s}\sum_{t=0}^{\infty}\sum_{t'=0}^{t}
\frac{s^{t+1}}{t+1}P_n^{(0)}(t')
=\frac{q}{s}\sum_{t=1}^{\infty}\sum_{t'=0}^{t-1}\frac{s^{t}}{t}P_n^{(0)}(t')
=\frac{q}{s}\sum_{t'=0}^{\infty}\left(\sum_{t=t'+1}^{\infty}\frac{s^{t}}{t}\right)P_n^{(0)}(t').
\end{equation}
Denoting $S=\sum_{t=t'+1}^{\infty}\frac{s^{t}}{t}$ with $|s|<1$, we have $\frac{\partial S}{\partial s}= \sum_{t=t'+1}^{\infty}s^{t-1}=s^{t'}/(1-s)$. Since $S(s=0,t')=0$, we deduce,
\begin{equation}\label{appS}
\sum_{t=t'+1}^{\infty}\frac{s^{t}}{t}=\int_{0}^{s}du \frac{u^{t'}}{1-u}.
\end{equation}
Inserting Eq. (\ref{appS}) into (\ref{app3rd}) and combining with (\ref{appLHS}), we obtain the $s$-transform of Eq. (\ref{appmaster}):
\begin{equation}
{\tilde P_n}^{(0)}(s)=\delta_{n,n_0}+\frac{1-q}{2}s
[{\tilde P_{n+1}}^{(0)}(s)+ {\tilde P_{n-1}}^{(0)}(s)]
+q\int_0^{s}du \frac{{\tilde P_n}^{(0)}(u)}{1-u}.
\end{equation}
We apply the Fourier transform (\ref{fourierlaplace}) to this equation and obtain
\begin{equation}
(1-b_k s) \widehat{\tilde P}^{(0)}(k,s)= e^{-ikn_0} + q \, \int_0^s \frac{\widehat{\tilde P}^{(0)}(k,u)}{1-u}\, du
\label{fk.3}
\end{equation}
where $b_k= (1-q)\, \cos(k)$. Taking the derivative with respect to $s$ and rearranging terms leads to a first order differential equation:
\begin{equation}
\frac{d\widehat{\tilde P}^{(0)}(k,s)}{ds} = 
\left[\frac{b_k(1-\alpha_k)}{1-bks}+\frac{\alpha_k}{1-s}\right]
\widehat{\tilde P}^{(0)}(k,s)
\quad {\rm 
where}\quad \alpha_k= \frac{q}{1-b_k},
\end{equation}
which is easily solved as
\begin{equation}
\widehat{\tilde P}^{(0)}(k,s) =e^{-ikn_0} (1-s)^{-\alpha_k}\,(1-b_k\,s)^{-(1-\alpha_k)},
\label{fk.4}
\end{equation}
after applying the condition $\widehat{\tilde P}^{(0)}(k,s=0)=e^{-ikn_0}$.
Expanding in powers of $s$, one can then write $\widehat{P}^{(0)}(k,t)$ explicitly as
\begin{equation}
\widehat{P}^{(0)}(k,t)= e^{-ikn_0}\sum_{m=0}^t \frac{(\alpha_k)_m\, (1-\alpha_k)_{n-m}}{m!\, 
(n-m)!}\, b_k^{n-m}\, , 
\label{fkt.1}
\end{equation}
where $(a)_n=a(a+1)...(a+n-1)$ and $(a)_0=1$.

%
%
%

\section*{References}

\providecommand{\newblock}{}

\end{document}